\documentclass[prb, twocolumn]{revtex4-2}

\usepackage{graphicx}
\usepackage{subcaption}
\usepackage{dcolumn}
\usepackage{bm}
\usepackage{hyperref}
\usepackage{xcolor}
\usepackage{mathrsfs}
\usepackage{enumitem} 
\usepackage{placeins}
\usepackage{academicons}
\usepackage{orcidlink} 
\usepackage{amsmath}
\usepackage{amsfonts}
\usepackage{tcolorbox}
\usepackage{soul}
\newcommand{\ba}{\begin{eqnarray}}
\newcommand{\ea}{\end{eqnarray}}

\usepackage[english]{babel}

\newcommand{\ignore}[1]{}

\usepackage{ragged2e}  

\begin{document}

\title{
Stabilizer Entropy and entanglement complexity in the Sachdev-Ye-Kitaev model
}

\author{Barbara Jasser\orcidlink{0009-0004-5657-2806}$^{1,2}$}
\email{b.jasser@ssmeridionale.it}
\author{Jovan Odavi\'{c}\orcidlink{0000-0003-2729-8284}$^{2,3}$}
\email{jovan.odavic@unina.it }
\author{Alioscia Hamma\orcidlink{0000-0003-0662-719X}$^{1,2,3}$}
\email{alioscia.hamma@unina.it }

\affiliation{$^1$Scuola Superiore Meridionale, Largo S. Marcellino 10, 80138 Napoli, Italy}
\affiliation{$^2$Istituto Nazionale di Fisica Nucleare (INFN), Sezione di Napoli, Italy}
\affiliation{$^3$Dipartimento di Fisica `Ettore Pancini', Universit\`a degli Studi di Napoli Federico II, Via Cintia 80126, Napoli, Italy}

\begin{abstract}
The Sachdev-Ye-Kitaev (SYK) model is of paramount importance for the understanding of both strange metals and a microscopic theory of two-dimensional gravity. We study the interplay between Stabilizer R\'enyi Entropy (SRE) and entanglement entropy in both the ground state and highly excited states of the SYK-4+SYK-2 model, interpolating the highly chaotic four-body interactions model with the integrable two-body interactions one. The interplay between these quantities is also assessed through universal statistics of the entanglement spectrum and its anti-flatness. We find that SYK-4 is indeed characterized by a complex pattern of both entanglement and non-stabilizer resources, while SYK-2 is non-universal and not complex. 
We discuss the fragility and robustness of these features depending on the interpolation parameter. 
\end{abstract}

\maketitle

\section{Introduction}

The Sachdev-Ye-Kitaev (SYK) model describes the behavior of strongly correlated fermions in strange metals, and has received significant attention from the high-energy community due to its profound connection with black hole physics~\cite{sachdev1993gapless, KITP2015, chowdhury2022sachdev}.
 The low-energy sector of this model is described by Jackiw-Teitelboim (JT) gravity, providing insight into the holographic description of black holes and their thermodynamic properties~\cite{sarosi2017ads,trunin2021pedagogical}. In the SYK model, the out-of-time-ordered correlators exhibit exponential growth over time with a growth rate that reaches the universal upper bound established in \cite{maldacena2016bound} and expected in a theory of gravity~\cite{Maldacena_Stanford_2016, Shenker_Stanford_2014}. Beyond this exciting connection to black hole physics, the SYK model has also been studied for its potential to be used as a quantum battery \cite{RevModPhys.96.031001, PhysRevResearch.2.023113, PhysRevB.98.205423,Sisorio_Cappellaro_DellAnna_2025}, exhibiting super extensive charging power \cite{PhysRevLett.125.236402}.

The most general variant of the model, the SYK-q, includes q-body interactions between Majorana modes, and the interpolation from the four-body SYK-4 to the two-body SYK-2 model offers a physically motivated route to understanding strongly correlated electron systems~\cite{PhysRevLett.125.196602, garcia2018chaotic}. The two theories are different: the SYK-4 is chaotic, with an exponential density of states at low energy, while the SYK-2 is a free random theory, with a polynomially vanishing gap~\cite{chowdhury2022sachdev, sarosi2017ads}. Remarkably, both models exhibit a volume law for entanglement~\cite{liu2018quantum, Huang_Tan_Yao_2024, fu2016numerical, zhang2022quantum}, prompting the question of what else sets these two models apart.

In quantum information theory, it is known that beyond entanglement, the next layer of quantum complexity is captured by \emph{non-stabilizerness}~\cite{Veitch_2014}, a fundamental resource for universal quantum computation, quantum error correction~\cite{shor1995scheme, calderbank1996good, bennett1996mixed, knill1997theory, gottesman1997stabilizer}, and quantum simulation~\cite{shor1996proceedings, gottesman1998theory, kitaev2003fault, campbell2017roads}. Recently, non-stabilizerness has risen to prominence due to finding a unique computable monotone for pure states, the Stabilizer R\'enyi Entropy (SRE)~\cite{leone2022stabilizer}. In the context of high-energy physics, both SRE and entanglement are studied for heavy nuclei simulations~\cite{robin2024magic} and neutrino physics~\cite{brokemeier2024quantum, chernyshev2024quantum}.  Moreover, the delocalization due to the entanglement of non-stablizerness resources has recently been connected to the holographic dual of back-reaction in the context of AdS-CFT  \cite{cao2024gravitationalbackreactionmagical}, features of CFT  \cite{Oliviero_Leone_Hamma_2022, White_Cao_Swingle_2021}, and the harvesting of quantum resources from the vacuum of a quantum field \cite{nystrom2024harvesting, cepollaro2024harvesting}.  When entanglement delocalizes and scrambles non-stabilizer resources, it gives rise to universal behavior  (expected from states sampled according to the Haar measure) of out-of-time-ordered correlation functions~\cite{leone2022stabilizer}, entanglement fluctuations \cite{Leone2021quantumchaosis, oliviero_transitions_2021} and the onset of chaotic behavior in quantum many-body systems \cite{yang2015two, Zhou_Yang_Hamma_Chamon_2020, Russomanno_Passarelli_Rossini_Lucignano_2025} in the ETH-MBL transition \cite{Yang_Hamma_Giampaolo_Mucciolo_Chamon_2017, rigol2008thermalization, pal2010many, nandkishore2015many}.  On the other hand, the fine structure of entanglement revealed by the statistics of entanglement gaps (ESS)  \cite{Chamon_Hamma_Mucciolo_2014, Shaffer_Chamon_Hamma_Mucciolo_2014, Zhou_Yang_Hamma_Chamon_2020} reaches the complex universal patterns of random matrix theory thanks to non-stabilizerness. 

\begin{figure}[t!]
    \centering  
    \includegraphics[width=\columnwidth]{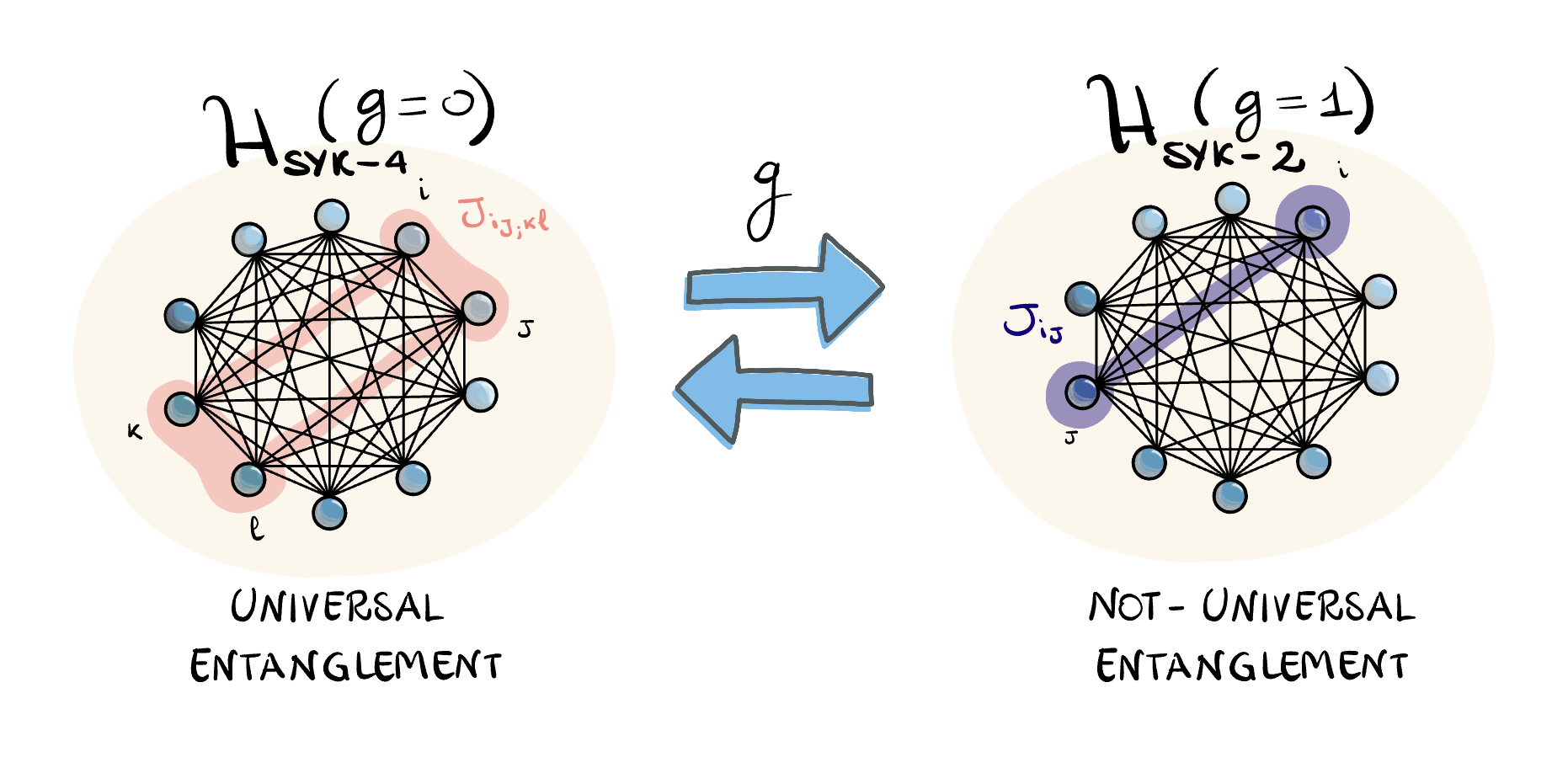}
    \caption{\justifying Schematic representation of the interpolated SYK-4 + SYK-2 model defined in Eq.~(5). At \(g=0\), the model reduces to SYK-4, a maximally chaotic system characterized by universal entanglement properties. At \(g=1\), the model reaches the SYK-2 limit which behaves like a disorder free-fermion system. In the ground state, SYK-4 features are fragile under finite values of the interpolation parameter \(g\), whereas those of SYK-2 remain robust. By contrast, SYK-4 behavior persists at high energies for all \(g \ne 1\), while making SYK-2 regime fragile.}
    \label{Hint}
\end{figure}

The main goal of this work is to study the SYK-4+SYK-2 model beyond the perturbative regime under the lens of the emergence of quantum complex behavior resulting from the interplay between entanglement and SRE.   We find that SYK-4 exhibits a non-trivial interplay of entanglement and SRE both in the ground state and in the high-energy eigenstates (middle-of-the-spectrum states), while SYK-2 shows features that are typical of integrable or non-chaotic models~\cite{Lau_Ma_Murugan_Tezuka_2019, lau2021correlated}. This is revealed by the adherence of SYK-4 to Haar-like behavior for entanglement entropy, reduced density matrix eigenvalues, and gaps statistics, capacity of entanglement, higher values of SRE, especially in its non-local features (see Fig.~\ref{Hint}). In contrast, SYK-2 exhibits lower SRE and displays non-universal behavior across all the aforementioned figures of merit.

We address the robustness of universal features in the interpolated model, where the parameter \( g \) controls the relative contribution of the SYK-4 and SYK-2 components. As defined in Eq.~\eqref{Hinterpolated}, the model reduces to SYK-4 when \( g = 0 \), while only the SYK-2 term remains at \( g = 1 \). Our results reveal that the universal properties of SYK-4 are highly fragile in the ground state: even for arbitrarily small \( g > 0 \), the system transitions into the SYK-2 class. At high energies, however, this behavior is reversed—the SYK-4 phase remains robust and persists across the entire range \( g < 1 \).

To further probe this transition, we analyze the entanglement spectrum and find that its adherence to Wigner-Dyson statistics is extremely sensitive to the interpolation, breaking down for both ground and high-energy states. To quantify this fragility, we employ tools inspired by quantum information theory, including the Kullback-Leibler (KL) divergence and a novel similarity measure we introduce, termed \emph{KL fidelity}. These insights not only complement but also extend the findings of Ref.~\cite{PhysRevLett.125.196602}, which were based on Green's function analysis. More specifically, we find that the critical interpolation parameter exhibits a power-law scaling, $g_c \sim N^{-3/4}$, and therefore vanishes in the thermodynamic limit. Thus, as $N \to \infty$, this shows quantitatively how the SYK-4 character of the ground state becomes unstable for arbitrarily small quadratic perturbations. Finally, we show that the behavior of SRE in the eigenstates can distinguish among the $8$-fold symmetry classes of SYK-4. This is remarkable, as such classification is typically associated with properties of the full Hamiltonian and its eigenspectrum, rather than individual eigenstates.

The paper is structured as follows.  We begin by introducing the SYK-q model and its interpolating version in Section~\ref{sec:model}, detailing the numerical methods used for our analysis. We then study the behavior of entanglement and its spectrum across the interpolation in Sections~\ref{sec:entanglement} and~\ref{sec:entanglementspectrum}, focusing on differences between ground and highly excited states. Next, we explore entanglement spectrum statistics in Section~\ref{sec:ESS} and the role of non-stabilizer resources, quantified by the Stabilizer Rényi Entropy in Section~\ref{sec:SRE}. Finally, we investigate measures of anti-flatness in Section~\ref{sec:AF} and discuss how they capture the non-local structure of entanglement and magic in the model. In Section~\ref{sec:conclusion}, we provide a comprehensive summary of the findings this work and mention future perspectives.

\section{The model}\label{sec:model}
The most general form of SYK models considers a q-body all-to-all interaction between $N$ Majorana fermionic modes with the Hamiltonian 
\begin{equation}
H^{\mathrm{SYK}-q} = (i)^{q/2} \sum\limits_{1 \le i_{1} < \hdots <  i_{q} \le N} J_{i_1 i_2 \hdots i_{q}} \chi_{i_1} \chi_{i_2} \hdots \chi_{i_q},
\end{equation}
with $q$ an even integer number~\cite{sachdev1993gapless, KITP2015, chowdhury2022sachdev}. The disorder in the model is due to the couplings $J_{i_1, i_2, \dots, i_q}$ which are identical, independent distributed $\rm (i.i.d.)$ gaussian variables with vanishing mean and variance
\begin{equation}
\overline{J_{i_1, i_2, \dots, i_q}} = 0 \; ,  \; \; \; \overline{J_{i_1, i_2, \dots, i_q}^2} = \frac{(q-1)! J}{N^{q-1}}  \;    . 
\label{variance}
\end{equation}
For $q > 2$, the system exhibits quantum chaos~\cite{cotler2017black, garcia2018chaotic, orman2024quantum}, as evidenced by several established indicators. One key probe is level repulsion in the energy spectra, characterized by the statistical distribution of energy level spacings~\cite{adhikari2020limiting, forrester2006quantum, forrester2020many, zyczkowski2000truncations}. This indicates that the eigenstates of the Hamiltonian are highly delocalized and correlated, a hallmark of quantum chaos. The SYK model is defined on a complete graph, so the interactions are strongly non-local. Using the Jordan-Wigner transformations, it is possible to map the Majorana operators into Pauli spin strings, used in the numerical simulations \cite{bravyi2002fermionic}. Notice that each spin operator can be expressed in terms of two Majorana operators. Therefore, the total number of Majorana operators is double the number of Paulis. The simplest version of the SYK model is 
\ba
H^{\mathrm{SYK}-2} = i \!\!\!\! \sum\limits_{1 \le i < j \le N} \!\!\!\!\! J_{i,j} \chi_{i} \chi_{j} \;. \label{modelSYK2}
\ea
Since the interactions between fermions are considered in pairs (see Fig.~\ref{Hint}), SYK-2 represents the free fermions point of the theory. This results in Gaussian statistics for its spectral properties~\cite{sarosi2017ads, Lau_Ma_Murugan_Tezuka_2019, lau2021correlated}, meaning that it does not show the same level of randomness and complexity found in many-body quantum chaotic systems, and makes it an example of a disordered fermionic system and which is analytically tractable \cite{liu2018quantum}. The four-body interaction model reads
\ba
H^{\mathrm{SYK}-4} = - \!\!\!\!\!\!\!\!\!\! \sum\limits_{1 \le i < j < k < l \le N} \!\!\!\!\!\!\!\!\!\! J_{i,j,k,l} \chi_{i} \chi_{j} \chi_{k} \chi_{l}  \; .
\label{modelSYK4}
\ea
The spectrum of the SYK-4 model exhibits clear signatures of quantum chaos, most notably level repulsion in the Hamiltonian eigenvalues~\cite{cotler2017black, garcia2018chaotic, orman2024quantum, kobrin2021many}. We define the interpolated the SYK-4 + SYK-2 model, as
\ba
H (g) := (1 - g) H^{\mathrm{SYK}-4} + g H^{\mathrm{SYK}-2}, \label{Hinterpolated}
\ea
with $g \in [0,1]$~\cite{PhysRevLett.125.196602}. In the following, we study the interpolated model using exact numerical diagonalization, focusing on the ground state (GS) and a middle-spectrum eigenstate (MS) across many disorder realizations \( M \). 

{From the point of view of experimental realizations, synthetic quantum matter offers the most promising route to realize interpolated and other SYK-type Hamiltonians, with both analog and digital strategies actively explored in the literature. Analog proposals range from solid-state mesoscopic systems~\cite{Pikulin_Franz_2017, Chew_Essin_Alicea_2017, Chen_Ilan_deJuan_Pikulin_Franz_2018, Luo_You_Li_Jian_Lu_Xu_Zeng_Laflamme_2019, Brzezińska_Guan_Yazyev_Sachdev_Kruchkov_2023, PhysRevB.108.064202}  to cold atoms in optical lattices~\cite{Danshita_Hanada_Tezuka_2017, Wei_Sedrakyan_2021, Uhrich_2023, Uhrich_Bandyopadhyay_Sauerwein_Sonner_Brantut_Hauke_2023, Sauerwein_Orsi_Uhrich_Bandyopadhyay_Mattiotti_Cantat-Moltrecht_Pupillo_Hauke_Brantut_2023}. In particular, the authors of~\cite{Brzezińska_Guan_Yazyev_Sachdev_Kruchkov_2023} were able to controllably engineer and enhance the SYK interactions in a disordered graphene flake using experimentally achievable magnetic fields. They achieve this result by simulating the effects of two sources of disorder: irregularities of system boundaries and bulk vacancies. On the other hand, some digital approaches based on gate-model quantum computation have also been proposed ~\cite{Babbush_Berry_Neven_2019, Asaduzzaman_Jha_Sambasivam_2024, Jha_2025, Granet_Kikuchi_Dreyer_Rinaldi_2025}. Together, these works mark significant progress toward realizing the SYK model in quantum simulators, even if the list is not exhaustive.
}

\section{Results}\label{sec:results}
\subsection{Entanglement}\label{sec:entanglement}
To quantify the bipartite entanglement, we focus on R\'{e}nyi entropies defined as
\ba
    S_{\alpha} = \frac{1}{1 - \alpha} \log{\mathrm{Tr} \left[ \rho^{\alpha}_{\mathrm{R}} \right]}, \quad \alpha \in [0, 1) \cup (1, \infty), \label{entropydef}
\ea
where the Von Neumann entanglement entropy 
\begin{equation}
    S_{1} (\rho_{\mathrm{R}}) = - \mathrm{Tr}(\rho_{\mathrm{R}} \log(\rho_{\mathrm{R}})) = -\sum\limits_{i} \lambda_i \log{ \lambda_i} \label{eq-VN},
\end{equation}
is recovered in the limit $\alpha \to 1^{+}$ and where $\lambda_i$ are eigenvalues of $\rho_{R}$. For random quadratic Hamiltonians, a class of models to which the SYK-2 Hamiltonian belongs~\cite{liu2018quantum}, the average entanglement entropy was derived in closed form in \cite{lydzba2020eigenstate}, reading
\begin{align}
    \mathcal{S}_{1}^{\mathrm{SYK-2}} \left(R, f\right) = \mathcal{K}(f) \ln{(2)} R,
\end{align}
where 
\begin{align}
    \mathcal{K}(f) = \left[1 - \frac{1 + f^{-1}(1-f) \ln{(1-f)}}{\ln{2}} \right].
\end{align}

\begin{figure}[t!]
   \centering
    \includegraphics[width=\columnwidth]{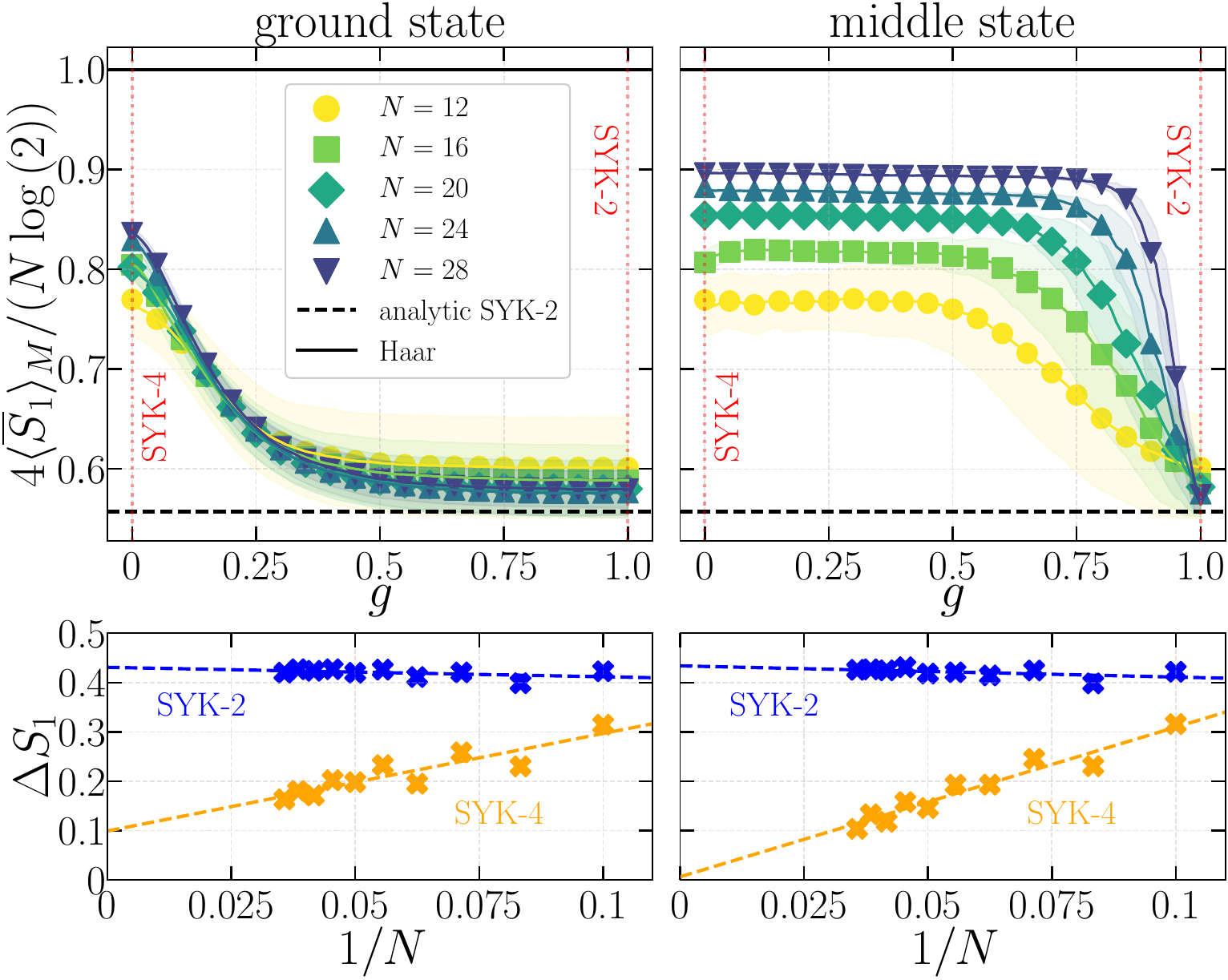}
    \caption{\justifying \textit{Upper panels}: Averaged bipartite von Neumann entanglement entropy, defined in Eq.~\eqref{eq-VN}, for the SYK-4 + SYK-2 model as a function of the interpolation parameter~$g$. Shaded regions indicate the standard deviation over $M$ disorder realizations. \textit{Lower panels}: Finite-size scaling of the relative entanglement entropy gap, defined in Eq.~\eqref{eq-relativegap}, along with its linear extrapolation to the thermodynamic limit. This figure highlights the fundamental difference in entanglement (i.e., non-local correlations) between the ground state (GS) and a middle-spectrum eigenstate (MS) across the interpolation. The left column shows results for the GS, while the right column corresponds to a MS.
    }\label{fig-entanglement}
\end{figure}

Here, $f = R / N^{\rm spin}$ represents the ratio between the subsystem size $R$ and the total system size in terms of the number of qubits. The entanglement entropy scaling of the SYK-2 eigenstates follows a volume law (extensive scaling with subsystem size) but with a coefficient dependent on the subsystem-to-system size ratio $f$, distinguishing it from the fully quantum chaotic regime in the thermodynamic limit. Efforts to extend these insights to the SYK-4 model include works such as Refs.~\cite{Huang_Gu_2019, Zhang_Liu_Chen_2020, Huang_Tan_Yao_2024}. 

The maximally chaotic random pure state, sampled uniformly according to the Haar measure, exhibits the Page value for entanglement entropy, given by
\begin{align}
    \frac{2 S_{1}^{\mathrm{Haar}}}{N \ln{(2)}} = 2f, \quad \mathrm{for} \quad f \in [0, 1/2],
\end{align}
to leading order in the system size~\cite{Page_1993, Odavic_Viscardi_Hamma_2024}.

In upper row panels of Fig.~\ref{fig-entanglement}, we show the rescaled average entanglement entropy in the ground state (GS) and middle-of-the-spectrum (MS) eigenstates of the interpolated SYK Hamiltonian \(H(g)\), with \(f = 1/2\). The average is taken over subsystem bipartitions (denoted \(\overline{S}\)) and disorder realizations (denoted \(\langle \cdot \rangle_{M}\)) for each value of the interpolation parameter \(g\). Disorder statistics for various system sizes are detailed in Appendix~\ref{SMstatistics}.

 We note that due to the all-to-all connected nature of the underlying graph structure of the SYK-$q$ modes, the number of possible bipartitions grows factorially with system size, making it computationally infeasible to consider all of them when computing averages. However, the system is on average permutationally invariant so we do not need to consider all the possible bipartitions. To make the calculation numerically tractable, we instead sample $N$ bipartitions from the total of $\binom{N/2}{N/4}$ possible ones.

 For \(g = 0\) (SYK-4), the GS entanglement entropy approaches the Haar (universal) value more closely than at \(g = 1\) (SYK-2), though neither reaches it. This is quantified by the relative gap 
\begin{equation}
    \Delta S_{1} := \left\vert \dfrac{ S_{1} -  S_{1}^{\rm Haar}}{ S_{1}^{\rm Haar}} \right\vert, \label{eq-relativegap}
\end{equation}
whose finite-size scaling is shown in the lower panels. In the lower left panel of Fig.~\ref{fig-entanglement}, the blue and orange dashed lines represent linear extrapolations to the thermodynamic limit \(1/N \to 0\), where we observe a finite, non-zero value. In the lower panels we fitted the data to the function of the form \( f(x) = a + bx \).  For the $g=0$ case, the extrapolated values are \( a = 0.1 \pm 0.02 \) (GS) and \( a = 0.007 \pm 0.017 \) (MS), indicating a vanishing entanglement gap in the thermodynamic limit for the highly chaotic regime. In contrast, the MS entanglement shows that the SYK-4 model achieves near-perfect agreement with the Haar value, while SYK-2 does not.  The behavior across \(g\) indicates that universal features of MS states persist for all \(g \ne 1\), highlighting the robustness of SYK-4 features in the higher energy states. To summarize, while it is well-known that the entanglement in SYK-4 obeys volume law for all the states, including the GS, we find that the GS deviates from the Haar value, while the MS states actually do reach the universal Haar value. This is not in contradiction with \cite{Huang_Tan_Yao_2024} as we compute entanglement exactly at half bipartition.

\subsection{Entanglement Spectrum}\label{sec:entanglementspectrum}

\begin{figure}[t!]
    \centering
    \includegraphics[width=\columnwidth]{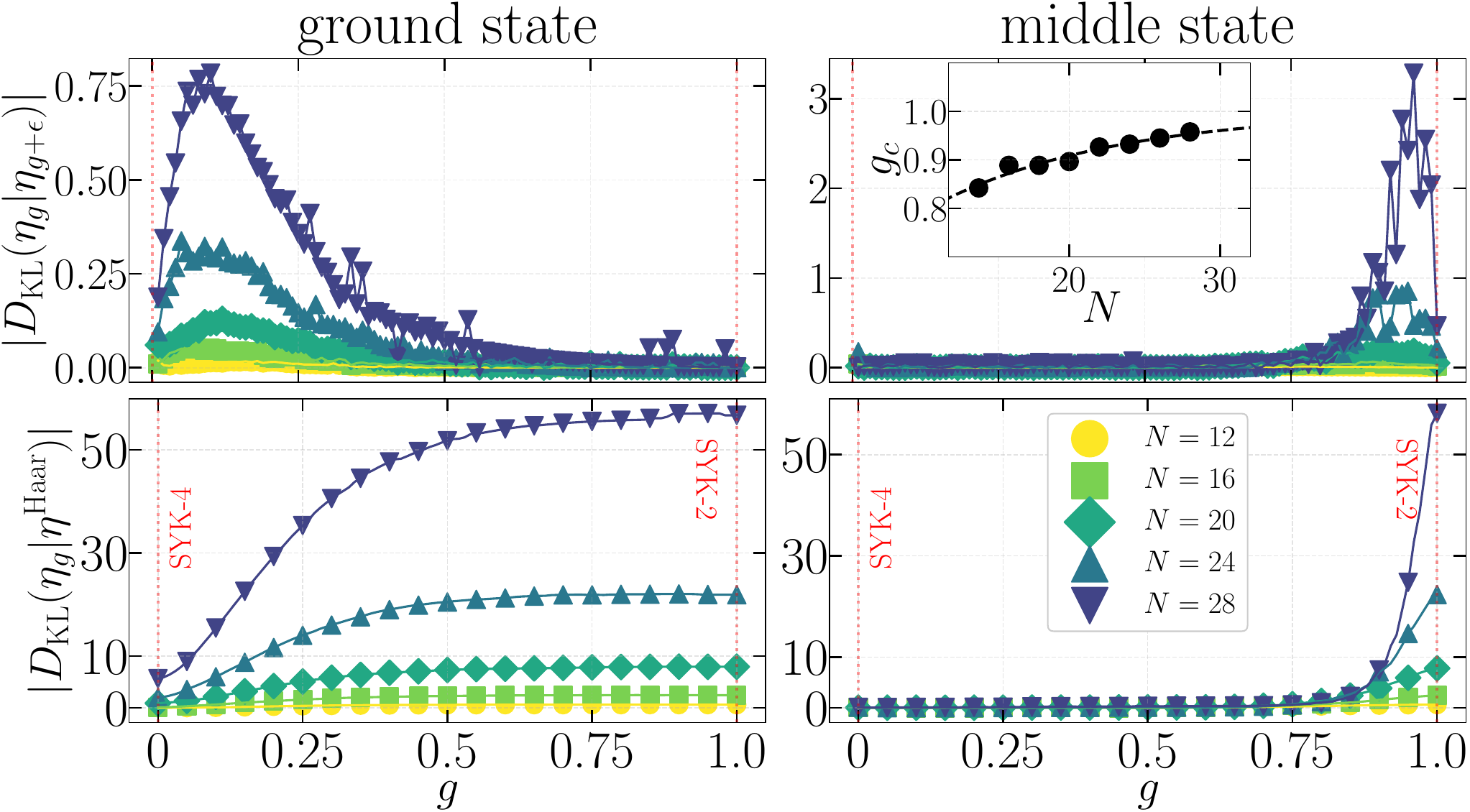}
    \caption{\justifying  Normalized reduced density matrix (RDM) eigenvalues of $f = 1/2$ subsystem-to-system. These are average values over many iterations~\ref{SMstatistics} across a single bipartition. One obtains similar results choosing other bipartitions. The reference value is the Marchenko-Pastur (M-P) distribution $\eta^{\rm Haar}(x)$ in Eq.~\eqref{etaHaar}. We define $\eta_{k} = k /d$ and $x_{k} = (1/2) \sqrt{ \lambda_{k} d}$ where $d = 2^{N/4}$ for $f=1/2$, while the binning step for the interpolation parameter has been set to $\epsilon = 0.01$. The left column shows results for the ground state (GS), while the right column corresponds to a middle-of-spectrum (MS) eigenstate. }\label{fig-normalizedRDM}
\end{figure}

A finer probe into the structure of entanglement is given by the full distribution of the eigenvalues of the reduced density Matrix (RDM) which we denote $\eta (x)$. For Haar-random pure states, this distribution corresponds to the Marchenko--Pastur (M-P) law, which arises as the limiting eigenvalue distribution of Wishart matrices~\cite{yang2015two, PhysRevD.109.126001}, and is given by 
\begin{equation}
\eta^{\rm Haar}(x) = 1 - \frac{2}{\pi} \left( x \sqrt{1 - x^2} + \arcsin{x} \right), \label{etaHaar}
\end{equation}
To distinguish two probability distributions we employ the  Kullback-Leibler divergence
\begin{equation}
D_{KL} (p||q):= \sum_i p_i(\log p_i - \log q_i).
\end{equation}
The results in the top row of Fig.~\ref{fig-normalizedRDM} show the  {\em KL fidelity} $D_{KL}(\eta_g|\eta_{g+\epsilon})$ between two distributions of the eigenvalues of the RDM for two nearby values $g, g+\epsilon$ of the interpolation parameter. This quantity can serve as a probe of a sharp transition associated with an observable consisting of a discrete probability distribution. We see that the GS and MS behave in a symmetric but opposite way. The structure of the eigenvalues of the RDM of the GS shows a sharp transition at $g>0$ and then smooths out. On the other hand, for the MS, $g=1$ is fragile, and the eigenvalues of the RDM of highly excited states are smoothly varying as long as SYK-4 interactions are different from zero. The second row of Fig.~\ref{fig-normalizedRDM} shows the same statistical distance between the state for the value $g$ and the reference Haar value. While SYK-4 is converging to the Haar value, SYK-2 shoots away. The robustness of the two phases for GS and MS, respectively, is confirmed. In other words, for every value of $g\ne 1$, the eigenvalues of the RDM of highly excited states behave very similarly to the universal Haar states.  The inset in Fig.~\ref{fig-normalizedRDM} shows the finite-size behavior of the peaks in the KL fidelity of the MS state from the numerical data. We find that the data are well described by an exponential of the form $f(N) = 1 - a e^{-b N}$ with $a = 0.49 \,\pm\, 0.08$ and $b = 0.084 \,\pm\, 0.009$ confirming the expected value of $g_c = 1$~\cite{garcia2018chaotic}.

\begin{figure}[t!]
    \centering
    \includegraphics[width=\columnwidth]{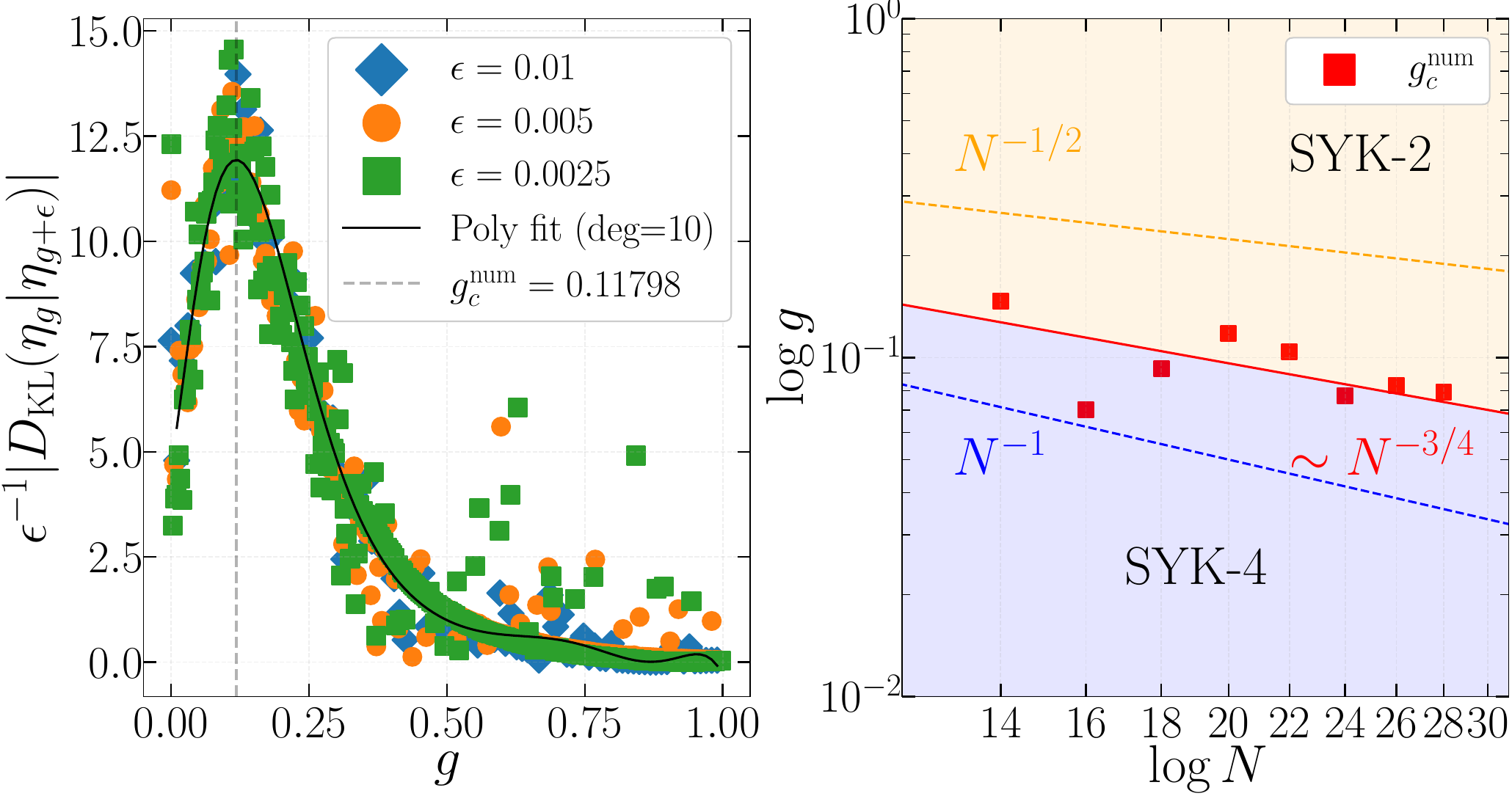}
    \caption{ \justifying Left panel: KL divergence–based fidelity as a function of the interpolation parameter \(g\) for different binning choices \(\epsilon\), with fixed system size \(N = 20 \). The rescaled fidelity diagnostic exhibits robust behavior across \(\epsilon\), justifying the use of \(\epsilon = 0.01\) to extract the transition point. The solid black line represents a degree-10 polynomial fit used to locate the peak (marked by the dashed vertical line). Right panel: Finite-size scaling of the extracted transition point \(g_c\), obtained from the location of the fidelity maximum. The fit reveals a power-law scaling, indicating that the SYK-4 ground state becomes unstable under arbitrarily small quadratic perturbations in the thermodynamic limit. In between $N^{-1}$ and $N^{-1/2}$ is where Ref.~\cite{PhysRevLett.125.196602} identified the potential transition to be located.   }\label{fig-transition}
\end{figure}

In the left panel of Fig.~\ref{fig-transition}, we show how the KL divergence–based fidelity behaves under different choices of the interpolation binning parameter \(\epsilon\), for a fixed number of Majorana fermions. We find that the rescaled fidelity diagnostic introduced in this work exhibits a consistent response across various binning choices. As a byproduct, this robustness provides justification for the specific choice of \(\epsilon = 0.01\) used in Fig.~\ref{fig-normalizedRDM}. Moreover, the stability of the response allows us to reliably use this binning to extract the location of the suspected transition point. To do so, we fit the data to a high-degree polynomial function, enabling a binning-independent estimate of the transition point (shown as the solid black curve). We found that a degree-10 polynomial captures the relevant feature, i.e. the location of the peak sufficiently well. The peak position, indicated by the dashed vertical line, marks the critical point for the transition.

In the right panel of Fig.~\ref{fig-transition}, we plot the transition point extracted from the polynomial fits as a function of system size \(N\). The resulting scaling behavior is well-described by the power law
\begin{equation}
    g_c = N^{-0.78 \pm 0.03} \sim N^{-3/4}.
\end{equation}
To the best of our knowledge, this is the first systematic numerical study that identifies this subtle crossover in the GS of the perturbed SYK model. Our analysis shows that the non-Fermi liquid GS of SYK-4 is unstable under SYK-2 type quadratic perturbations. Physically, this implies that the transition to Fermi liquid behavior occurs at a perturbation strength that vanishes in the thermodynamic limit, rendering the special SYK-4 character of the GS unstable for arbitrarily small quadratic terms as \(N \to \infty\).

Experimentally, measuring the KL-fidelity requires knowledge of the reduced density matrix (RDM) eigenvalues for different values of the interpolation strength $g$. In both analog and digital simulators, this amounts to extracting the RDM spectrum of a state for a chosen bipartition, which can be done via subsystem state tomography (measuring all Pauli string correlators)~\cite{Cramer_Plenio_Flammia_Somma_Gross_Bartlett_Landon-Cardinal_Poulin_Liu_2010}, entanglement spectroscopy (inferring eigenvalues from RDM moments)~\cite{Johri_Steiger_Troyer_2017}, or classical shadows and randomized measurements (reconstructing the RDM from local random measurements)~\cite{Huang_Kueng_Preskill_2020, doi:10.1126/science.aau4963}. Among these, shadow-based and randomized protocols are particularly promising for scalability, since the measurement cost grows exponentially only with the subsystem size, and remains largely independent of the full system size.

\subsection{ Entanglement Spectrum Statistics (ESS)}\label{sec:ESS}
Complex pattern of entanglement is characterized by universal properties of the statistics of the gaps in the entanglement spectrum, the so-called {\em entanglement spectrum statistics} (ESS) \cite{Chamon_Hamma_Mucciolo_2014}. Chaotic systems, such as non-integrable quantum systems obeying the eigenstate thermalization hypothesis (ETH), feature a universal, Wigner-Dyson (WD)  behavior for the ESS; while integrable, disordered free-fermion models feature a Poisson statistics in both the high energy states and the long time behavior away from equilibrium. Hybrid cases like  MBL systems feature deviations from WD in polynomial time after a quantum quench \cite{Yang_Hamma_Giampaolo_Mucciolo_Chamon_2017}. The transition between the two regimes is due to the injection of non-stabilizer resources, which are scrambled around~\cite{Zhou_Yang_Hamma_Chamon_2020, True2022transitionsin}. From the perspective of random matrix theory, if the eigenvalues of the reduced density matrix are uncorrelated, the distribution is Poissonian. Conversely, correlated eigenvalues follow the Wigner-Dyson universality class.

Here we focus on the probability density function (PDF) of the consecutive spacing ratios denoted as \( P(r) \). To evaluate it we use the ascending eigenvalues \(\{\lambda_k\}\) of the  RDM and determine the spacing ratio as  
\begin{equation}
r_k = \frac{\lambda_{k+1} - \lambda_k}{\lambda_k - \lambda_{k-1}}, \quad  k = 2, 3, \dots, 2^{R} - 1.
\end{equation}
The resulting ratios \(\{r_k\}\) are plotted as a normalized histogram, excluding rare outliers with \( r_j > 10.0 \) to ensure proper normalization and accurate binning~\cite{Odavic_Torre_Mijić_Davidović_Franchini_Giampaolo_2023, Odavic_Mali_2021}. The explicit functional forms of the corresponding Poisson and Wigner-Dyson Gaussian ensembles are provided in ~\cite{Atas_Bogomolny_Giraud_Roux_2013} and also Appendix~\ref{app:HamS}. To complement this type of analysis, we also evaluate the averaged consecutive spacing ratio, defined as 
\begin{equation}
\bar{r} = \Bigg\langle \Bigg\langle \frac{\min (s_{k,j},s_{k+1,j})}{\max(s_{k,j},s_{k+1,j})} \Bigg\rangle \Bigg\rangle_{2^{R}-2, M}. 
\end{equation}
where the spacings are given by \( s_{k,j} = \lambda_{k+1,j} - \lambda_{k,j} \), and the index \( j = 1, 2, \dots, M \) refers to the different RDMs considered across ensemble or disorder realizations.

\begin{figure}[t!]
\includegraphics[width=\columnwidth]{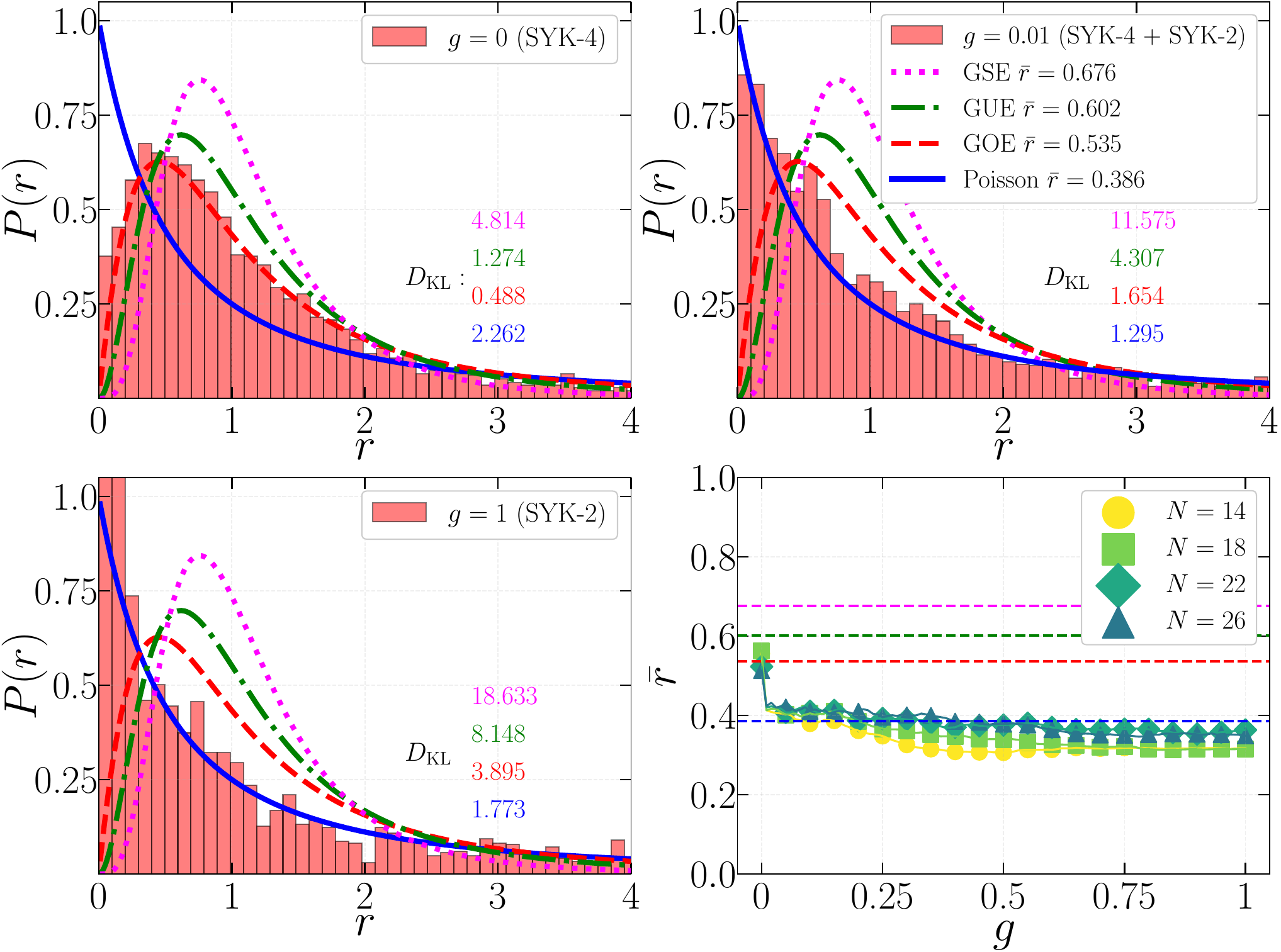}
    \caption{\justifying ESS of the GS RDM eigenvalues of the $H_g$ model for different values of $g$. We superimpose the analytical curves for the Wigner-Dyson (dashed) and Poisson (blue continuous) distribution for comparison. System size  $N = 22$ and number of realizations is $M = 100$. The number of bins used for the histogram is 100. The colored numbers represent the KL divergence of data against known distributions described in Appendix~\ref{app:HamS}. }\label{fig-ESS}
\end{figure}

\begin{figure*}[t!]
        \centering
\includegraphics[width=0.9\linewidth]{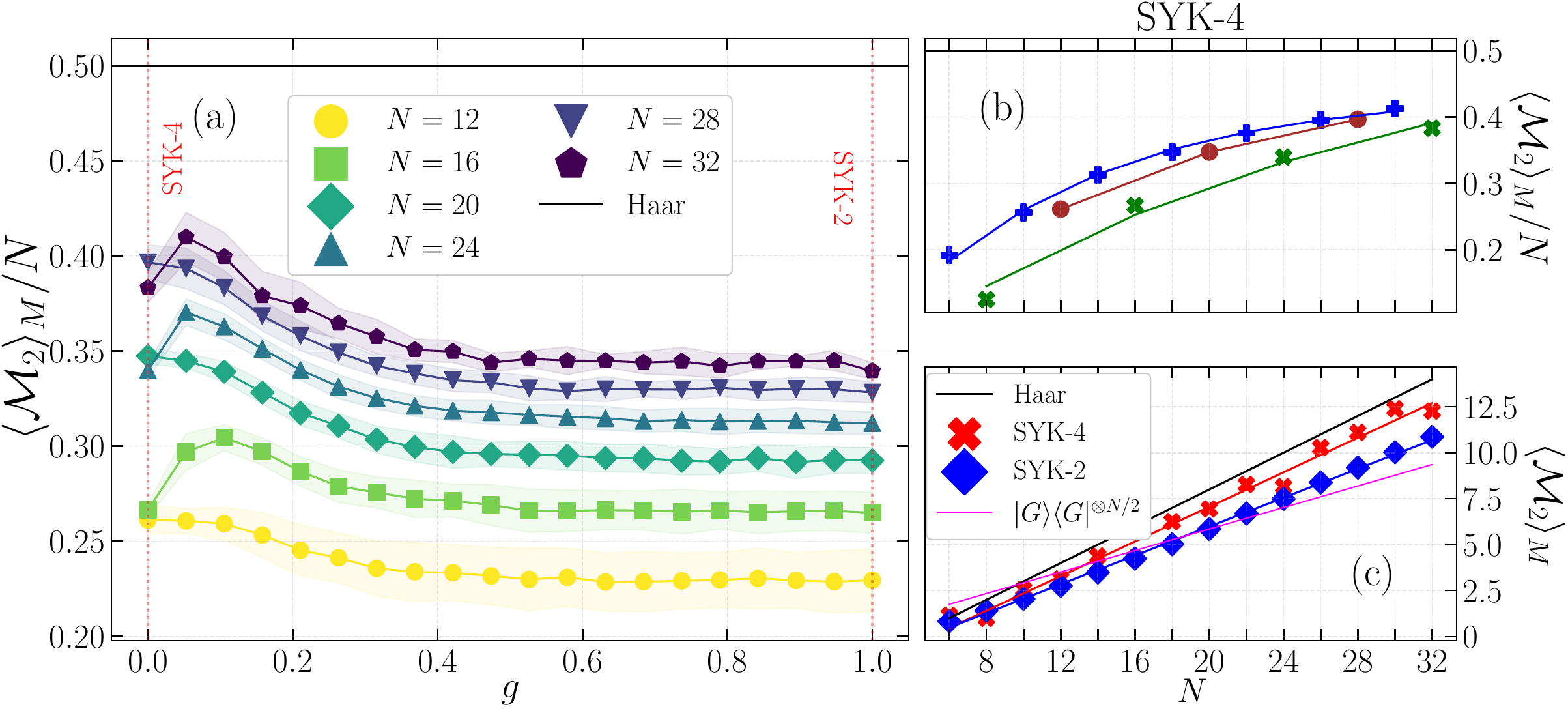}
    \caption{\justifying Ensemble averaged SRE $\mathcal{M}_{2}$ in the GS of the $H (g)$. \textit{Panel (a):}  For all $g$ the ground states exhibit a non-vanishing amount of non-stabilizerness. \textit{Panel (b)}: The non-monotonicity can be easily captured by a grouping of points: green crosses for $N\!\!\mod 8 = 0$, blue for $N\!\!\mod 8 = 2,6$, and red circles for $N\!\!\mod 8 = 4$. The lines represent fits to the data. More details on the obtained fitting parameters are provided in Appendix~\ref{app:SREe}. \textit{Panel (c)}: SRE without the normalization with the system size as compared to the Haar value. $|G\rangle \langle G|$ is a single qubit state which maximizes SRE (see Appendix~\ref{app:SREe}), indicating beyond local non-stabilizerness present in the model as $N \to \infty$.   }\label{fig-SRE}
\end{figure*}
In Fig.~\ref{fig-ESS}, we compute the ESS \( P(r) \) for different realizations of the GS of the $H(g)$ model for several values of $g$. 
The SYK-4 model adheres to the universal WD distribution for GOE. Once the model is perturbed by an SYK-2 term, the distribution shifts closer to a Poisson distribution. In the lower right panel of Fig.~\ref{fig-ESS}, we show the behavior of the averaged consecutive spacing ratio $\bar{r}$ as a function of $g$, which shows a sudden jump as one moves from $g=0$. We can see that this feature of the entanglement complexity, which is typical of chaotic systems, is fragile in the model. Similar results hold for the MS, see Appendix~\ref{app:MS}. 

\subsection{Stabilizer R\'{e}nyi Entropy}\label{sec:SRE}

Non-stabilizerness is an essential property  for universal quantum computation~\cite{Bravyi_Kitaev_2005}. To quantify this property, we used the so-called Stabilizer Rényi Entropy (SRE) $\mathcal{M}_{\alpha}$~\cite{leone2022stabilizer}, 
\begin{align}
    \mathcal{M}_{\alpha} (\Psi) = \dfrac{1}{1 - \alpha} \log_{2}{\left( d^{-1} \sum_{P \in \mathcal{P}_{N}} \vert {\rm Tr} (P \Psi) \vert^{2 \alpha}\right)}, \label{SRE}
\end{align}
the unique computable measure of non-stabilizerness~\cite{Leone_Bittel_2024}. The SRE can be efficiently evaluated without any minimization procedure. In Eq.~\ref{SRE}, $\Psi$ is the density operator of a $N$-qubit state $\vert \Psi \rangle$, $d = 2^N$ is the Hilbert space dimension, and $\mathcal{P}_N$ is the $N$-qubit Pauli group. The SREs for $\alpha \geq 2$ are good resource monotones~\cite{Leone_Bittel_2024}. Stabilizer entropies measure how far a state deviates from stabilizer states by analyzing its spread in the Pauli operator basis~\cite{leone2022stabilizer, Turkeshi_Tirrito_Sierant_2025S, Niroula_White_Wang_Johri_Zhu_Monroe_Noel_Gullans_2024}. Experimental protocols to measure this quantity have recently been proposed~\cite{Oliviero_Leone_Hamma_Lloyd_2022, PhysRevLett.132.240602}, allowing high-accuracy extraction of the SRE in digital quantum simulators.

In this work, we adopt the methodology outlined in Ref.~\cite{Odavic_Viscardi_Hamma_2024} to compute the SRE. Specifically, we transform the GS vectors obtained from exact diagonalization into matrix product state (MPS) tensor representations and employ the Perfect Sampling algorithm~\cite{Haug_Piroli_2023Quantifying, Lami_Collura_2023, Collura_Nardis_Alba_Lami_2025}. In Fig.~\ref{fig-SRE}, we show the numerical results for the scaling of SRE in the GS of the $H_g$ model. From panel $(a)$, we observe that the SYK-4 ($g = 0$) case exhibits a higher degree of non-stabilizer resources compared to the SYK-2 ($g = 1$) limit. From Fig.~\ref{fig-SRE} ($a$), we observe that the SRE of the SYK-4 GS does not exhibit a consistent monotonic increase with fermion number \(N\). A more detailed analysis in Fig.~\ref{fig-SRE} ($b$) reveals an oscillatory pattern in the SRE. The data is grouped and labeled according to the values of \(N\!\!\mod 8\), as indicated in the captions. Each of the three groups corresponds to a distinct exponent of the damped exponential. This grouping is motivated by the observation in~\cite{cotler2017black}, where it was first noted that the SYK-4 Hamiltonian exhibits a particular particle-hole symmetry. The authors linked this symmetry to different Gaussian random matrix universality classes: GOE for \(N\!\!\mod 8 = 0\); GUE for \(N\!\!\mod 8 = 2, 6\); and GSE for \(N\!\!\mod 8 = 4\), which manifest in the Hamiltonian's spectrum.  We observe the effects of this symmetry specifically in the GS, without considering the full Hamiltonian spectrum. We verified numerically that this symmetry holds also in the GS. In contrast to entanglement (see Fig.~\ref{fig-entanglement}) and ground state energy~\cite{García-García_Verbaarschot_2016}, the SRE is a probe of quantum complexity sensible to the finiteness of \(N\). This highlights the unique ability of SRE to reveal hidden structures within many-body systems. A recent, striking example is the ability of SRE to detect a quantum phase transition that is not captured by entanglement in a system without a conventional order parameter~\cite{Catalano_Odavic_Torre_Hamma_Franchini_Giampaolo_2024}.

How does the average SRE evolve as the system size increases in the GS and MS of the SYK-4 Hamiltonian? This behavior is depicted in panel $(c)$ of Fig.~\ref{fig-SRE} for the GS, and in the inset of Fig.~\ref{figMiddleSRE} in Appendix~\ref{app:SREMS} for the MS. By fitting the data to a linear function~\cite{Liu_Winter_2022}, we obtain the following expression for the average SRE:
\begin{eqnarray}
\mathcal{M}_{2}^{\rm SYK-4, GS} \sim - 2.4 + 0.95 \frac{N}{2} \; \ , \label{eq:res1}\\
 \mathcal{M}_{2}^{\rm SYK-4, MS} \sim - 2.6 + 0.96 \frac{N}{2} \; \ .\label{eq:res2}
\end{eqnarray}
For Haar random states, it is known that
\begin{align}
    \mathcal{M}_{2}^{\rm Haar} = - 2 + \dfrac{N}{2}, \label{eq: resH}
\end{align}
to leading order in the system size~\cite{leone2022stabilizer, Turkeshi_Dymarsky_Sierant_2025,Odavic_Viscardi_Hamma_2024}. The factor $2$ in the denominator of the linear term arises because \(N/2\) represents the support of the spin/qubit representation. Our results in Eq.~\eqref{eq:res1} and Eq.~\eqref{eq:res2}, as compared to Eq.~\eqref{eq: resH}, indicate that both the GS and MS of the SYK-4 model exhibit deviations from the behavior expected of fully quantum chaotic and universal states. Specifically, we observe a linear prefactor difference of approximately $0.05$ relative to Haar-random states. These observations support the conclusion that the eigenstates of the interpolated SYK-4 + SYK-2 model do not fully achieve universality.

\subsection{Entanglement Spectrum Anti-Flatness}\label{sec:AF}

A crucial insight in understanding the relationship between non-stabilizerness and entanglement resources comes from the fact that stabilizer states have a flat entanglement spectrum \cite{hamma2005bipartite}. Indeed, it is known that there is a strict relationship between the lack of flatness of the entanglement spectrum of a RDM and the non-stabilizerness of the full state. This is remarkable as anti-flatness is a local (that is, pertainind to the reduced density operator) quantity while SRE pertains to the full state. More specifically, it was shown~\cite{tirrito2024quantifying} that there is a proportionality relation between the linear 2-SRE (see Eq.~\eqref{SRE}), $\mathcal{M}_{2}^{\rm lin} (\Psi) = 1 - d^{-1} \sum_{P \in \mathcal{P}_{N}} \vert {\rm Tr} (\Psi P) \vert^{4}$, and a particular measure of the anti-flatness of a RDM.
This establishes an explicit connection between bipartite entanglement properties and the total system’s non-stabilizerness response. In particular, a non-flat entanglement spectrum implies that the state is both entangled with respect to the chosen bipartition and possesses non-stabilizer features. Thus, anti-flatness serves as a diagnostic of magic that is invariant under local unitaries acting within each subsystem of the bipartition. 

The non-local non-stabilizerness of a state is defined as the component of its non-stabilizerness that cannot be removed by local unitary operations~\cite{cao2024gravitationalbackreactionmagical,cepollaro2024harvesting}. Thus, it is natural to define the non-local SRE to measure this quantity. The non-local SRE~\cite{cao2024gravitationalbackreactionmagical} of a state $|\Psi\rangle$ is defined as $\mathcal{M}_{\alpha}^{NL}(|\Psi\rangle) = \underset{U_A \otimes U_B}{\min} \mathcal{M}_{\alpha}(U_A \otimes U_B|\Psi\rangle)$, where the minimization is over all local unitaries acting on the bipartition $A\otimes B$. A vanishing value of $\mathcal{M}_{\alpha}^{NL}(|\Psi\rangle)$ means that the non-stabilizerness of the state is entirely local, thus it can be eliminated by local unitaries. In this case, the entanglement spectrum of the reduced density matrix is flat, establishing a direct connection between non-local non-stabilizerness and the structure of bipartite entanglement.

One of the most useful measures of anti-flatness comes from the modular entropy~\cite{Dong_Lewkowycz_Rangamani_2016}
\begin{align}
   \tilde{S}_{\alpha} :=  \alpha^{2} \partial_{\alpha} \left( \dfrac{\alpha - 1}{\alpha} S_{\alpha} \right).
\end{align}
Its derivative, with respect to the R\'enyi parameter at $\alpha =1$, is minus the variance of the entanglement Hamiltonian $H_\rho := - \log \rho$,
\ba
- \operatorname{Var}_{\rho}(H_\rho) = - \langle\log^2\rho\rangle_\rho + \langle\log\rho\rangle_\rho^2
 =\partial_{\alpha} \tilde{S}_{\alpha} \Big\vert_{\alpha = 1}\ ;\label{CE}
\ea
also known as capacity of entanglement $C_E$, and it is able to quantify the anti-flatness of a spectrum~\cite{yao2010entanglement,schliemann2011entanglement,de2019aspects}.
For Haar random states, one can compute
\begin{equation}
   C_{E}^{\rm Haar} = \partial_{\alpha} \tilde{S}_{\alpha}^{\rm Haar} \Big\vert_{\alpha = 1} = \dfrac{11}{4} - \dfrac{\pi^2}{3} \approx -0.539868.
\end{equation}
which we use in our analysis as a reference value~\cite{de2019aspects, Okuyama_2021}. We note that  $C_{\rm E}$ for the SYK-2 has been obtained in~\cite{Bhattacharjee_Nandy_Pathak_2021}

\begin{figure}[t!]
    \centering
    \includegraphics[width=0.475\textwidth]{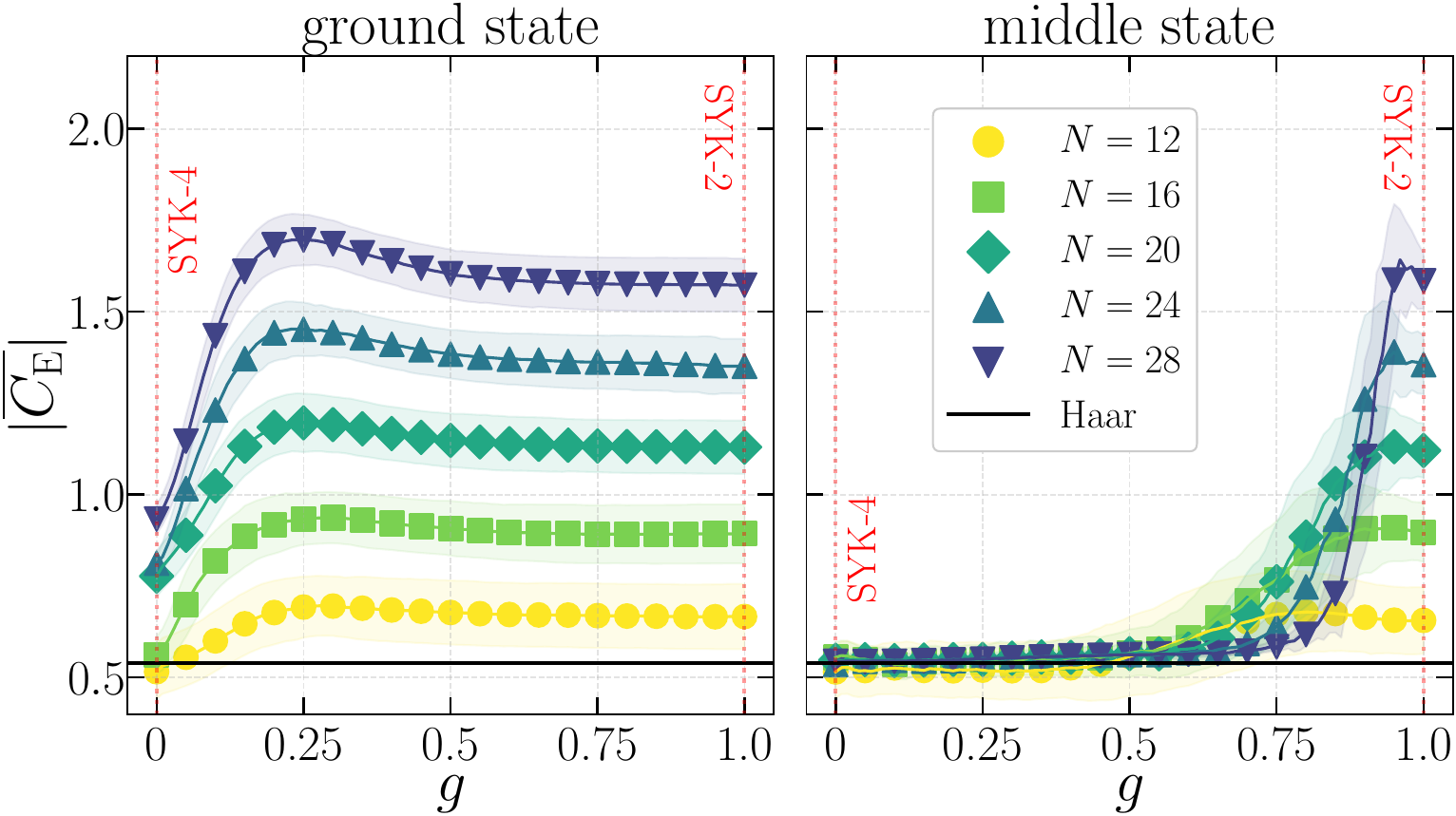}
    \caption{\justifying Absolute value of the averaged half-system capacity of entanglement $C_{E}(\rho_{\mathrm{R}})$ in the ground state  (left panel) and middle-of-the-spectrum states (right panel) of $H_g$.  
    }\label{fig-antiflatness}
\end{figure}

In Fig.\!~\ref{fig-antiflatness},  we compute the averaged capacity of entanglement $C_E$ as a function of $g$ in both the GS and MS state of $H (g)$. We observe that SYK-4 features a Haar-like capacity of entanglement, which is in a good agreement in the MS. Our study of the capacity of entanglement supports our claims on robustness/fragility observed with previous probes: while for the GS, the SYK-4 features are fragile, they are robust and can be extended for any $g<1$ in the MS. 

In Appendix~\ref{app:antiflatnessderivation}, we show the results for another measure of antiflatness, the logarithmic anti-flatness, expressed as the difference between R\'{e}nyi entropies. For this measure we obtain an analytical expression of it incthe GS of the SYK-2 Hamiltonian. We find that the general behavior of this quantity strongly follows the trends of the capacity of entanglement shown here. 

\section{Conclusions and Outlook}\label{sec:conclusion}

In this work, we explored the interplay between entanglement and non-stabilizer resources in the interpolating SYK-4+SYK-2 model using a range of diagnostic tools. These include measures such as entanglement entropy, stabilizer entropy, entanglement spectrum statistics, and their comparison with predictions from random matrix theory and Haar-random states~\cite{Turkeshi_Dymarsky_Sierant_2025}. We find that, while several properties, especially in the middle of the spectrum, closely resemble Haar-random behavior, significant deviations persist. In particular, the stabilizer entropy systematically is different compared to the Haar expectation when extrapolated to the thermodynamic limit, both for ground state and for typical high-energy eigenstates. In contrast, the capacity of entanglement for high-energy states in SYK-4 aligns well with the Haar predictions, revealing a nuanced and measure-dependent picture of quantum complexity.

By introducing a Kullback-Leibler divergence-based probe of the entanglement spectrum, we identified a complexity transition in the GS of the interpolated SYK model. We find that the critical interpolation parameter scales as $\sim N^{-3/4}$ (where we set the coupling constant $J = 1$), indicating the instability of the SYK-4 phase in the presence of arbitrarily small quadratic perturbations in the thermodynamic limit in the GS of the interpolated model. This result extends the findings of Ref.~\cite{PhysRevLett.125.196602} and demonstrates that this information-theoretic probe is significantly more sensitive than the other diagnostics we employed.

Altogether, our study contributes to the ongoing discussion on the extent to which finite-sized SYK-4 model captures features of quantum black holes and fast scramblers~\cite{Orman_Gharibyan_Preskill_2024}. An interesting future direction would be to study the behavior of stabilizer entropy and related diagnostics analytically in the double-scaled limit~\cite{cotler2017black, Berkooz_Isachenkov_Narovlansky_Torrents_2019}, where greater theoretical control may offer deeper insights into the structure of quantum complexity in the SYK model.

The quantum advantage of SYK-based quantum batteries may stem from the optimal exploitation of both entanglement~\cite{PhysRevE.87.042123} and non-stabilizer resources~\cite{PhysRevResearch.2.023095, Caravelli2021energystorage, PhysRevLett.125.040601}, highlighting a promising avenue for future research as well.

One of the most remarkable aspects of SYK-4 is its duality with JT gravity in the low-temperature regime~\cite{Maldacena_Stanford_2016}. While our results show that SYK-4 is fragile under two-body perturbations in the GS but remains robust at high energies, the question of its stability at low temperatures remains open—an issue we aim to address in future work.

{\em Acknowledgments.---}
Upon completion of this paper, we stumbled onto arXiv:2502.01582 which has a similar scope with our work. An early version of our work was presented at the poster session of the first workshop on many-body quantum magic   (MBQM2024), TII Abu-Dhabi in November 2024: \href{https://mbqm.tii.ae/posters.php}{\em Barbara Jasser, 
Chaos, Entanglement and Stabilizer Entropy in SYK model}. AH and JO acknowledge support from the PNRR MUR project PE0000023-NQSTI. AH acknowledges support from the PNRR MUR project CN 00000013-ICSC. This work has been funded by project code PIR01 00011
‘IBISCo’, PON 2014-2020, for all three entities (INFN,
UNINA and CNR). Additionally, acknowledge ISCRA for awarding this project access to the LEONARDO supercomputer, owned by the EuroHPC Joint Undertaking, hosted by CINECA (Italy) under the project ID: PQC - HP10CQQ3SR. We acknowledge stimulating conversations with P. Zanardi and Lorenzo Campos Venuti.

\bibliography{Bibliography.bib}
\bibliographystyle{apsrev4-2}

\appendix

\section{On the evaluation of Stabilizer R\'{e}nyi Entropy}\label{app:SREe}

\begin{figure}[h!]
    \centering
\includegraphics[width=\columnwidth]{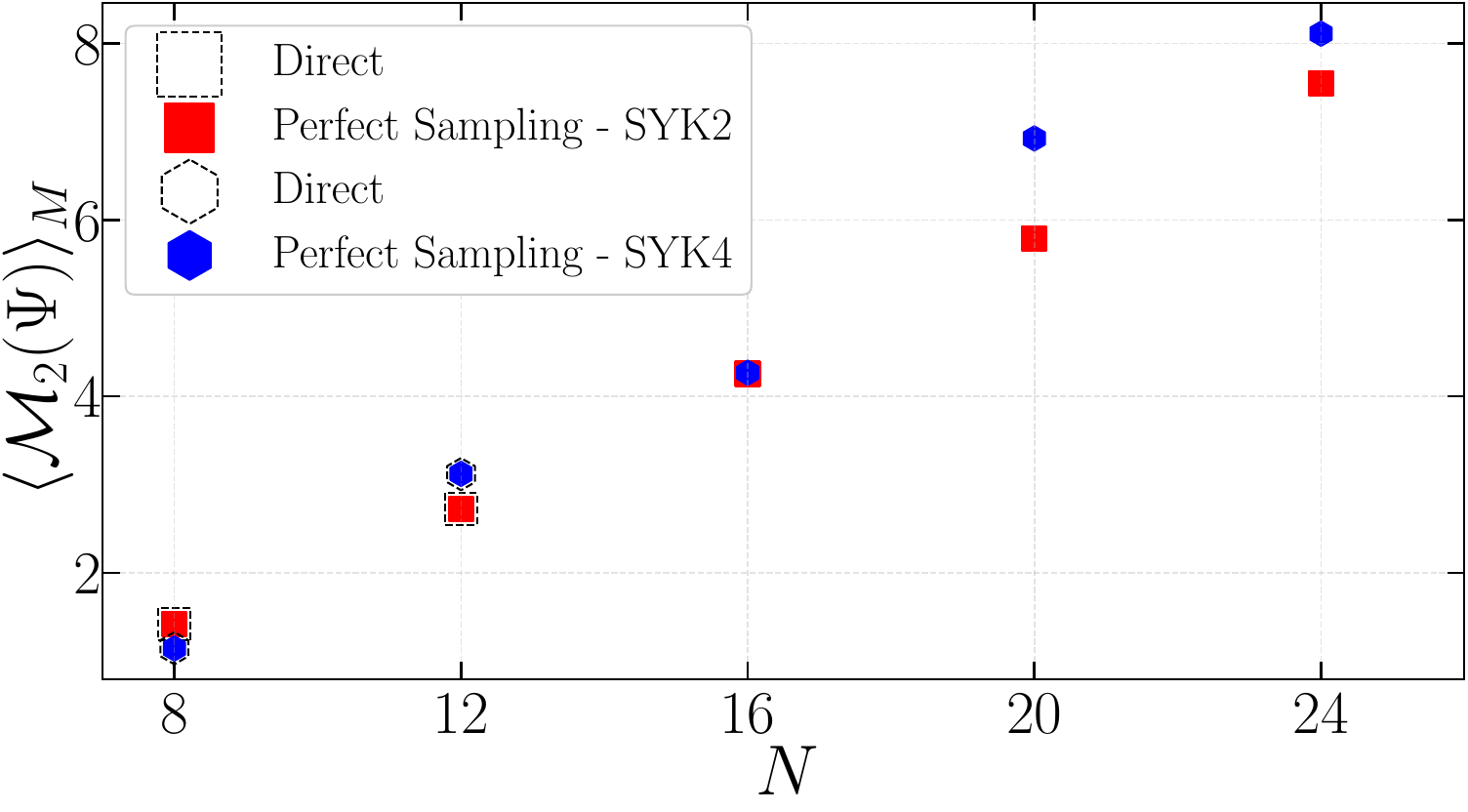}
    \caption{\justifying This figure shows the ability of the Perfect Sampling algorithm to evaluate the SRE efficiently. We use $M = 100$ different disorder realizations to compute the average SRE and perform an average (denoted in the $y$-label as $\langle \bullet \rangle_{M}$) over these realizations. We observe that for larger systems, the SYK-4 model host more magic as compared to the SYK-2 model. The coupling strength in general is set $J = 1$. }\label{fig-SREcheckSM}   
\end{figure}

Here, we outline key details of the Perfect Sampling algorithm~\cite{Haug_Piroli_2023Quantifying, Lami_Collura_2023}, essential for obtaining reliable estimates of the Stabilizer Rényi Entropy (SRE) presented in Fig.~(5). Starting from a state vector supported on \( N/2 \) qubits (where \( N \) denotes the number of Majorana fermions) obtained via exact diagonalization, we represent it as a Matrix Product State (MPS) with a maximum bond dimension of \( \chi = 2^{N/2} \) and a truncation cutoff of \( 10^{-8} \) using the ITensor library~\cite{10.21468/SciPostPhysCodeb.4,fishman_codebase_2022}. We then sample \( 10^{4} \) Pauli strings, leading to an absolute error in the stabilizer entropy estimation within the range \( 10^{-1} - 10^{-2} \).  

To verify the accuracy of this approach, we consider smaller system sizes in Fig.~\ref{fig-SREcheckSM}, where the SRE can be computed exactly from the state vector, evaluating all the $4^N$ expectation values of the Pauli strings, confirming that the method achieves sufficient precision. A similar approach was employed in~\cite{Odavic_Viscardi_Hamma_2024}. 

In Fig.~(5b) the green crosses are related to sizes such that $N\!\!\mod 8 = 0$, for which the fit parameters are $a = 0.57, b = 0.036$, blue crosses for $N\!\!\mod 8 = 2,6$, with $a = 0.44, b = 0.09$, and red circles for $N\!\!\mod 8 = 4$, with $a = 0.46, b = 0.07$. The lines represent fits to the function $f(x) = a (1 - \exp{(-b x)})$.  

In Fig.~(5c), we highlight a factorizable \( n \)-qubit state with the maximal amount of single-qubit magic, shown in magenta. It is known that there is a single qubit state that maximizes the SRE, the golden state $|G\rangle \langle G| = \frac{1}{2} \left( I + \frac{X+Y+Z}{\sqrt{3}} \right)$. The stabilizer entropy of its \( n \)-qubits product state is  $\mathcal{M}_{2} (|G\rangle \langle G|^{\otimes n}) = n \log_2 \left(\frac{3}{2}\right)$.

\section{Capacity of entanglement  - quick rederivation}
For the convenience of the reader, we compute here Eq.~(17) for the pure Haar random states. Original derivations were performed in~\cite{de2019aspects, Okuyama_2021}. For Haar random pure states, the average R\'{e}nyi entropy to leading order in $N$, with positive integer $\alpha$, is given by
\begin{align}
S_{\alpha} ^{\rm Haar}& \! = \! \frac{1}{1 - \alpha} \log{ \left[2^{N - R(1 + \alpha) } \sum_{k=1}^{\alpha} H(\alpha, k) 2^{(2 R - N)k}\right]} \label{Page}
\end{align}
where the coefficients $H(\alpha, k) = \frac{1}{\alpha} \binom{\alpha}{k} \binom{\alpha}{k-1}$ are known as Narayana numbers.
To start the calculation we simplify the expression first by fixing the system-to-subsystem size ratio to $R/N = 1/2$ to obtain for Eq.~\eqref{Page} the following
\begin{align}
    S_{\alpha}^{\rm Haar} = \dfrac{1}{1 - \alpha} \log{\left[ 2^{\frac{N}{2} (1 - \alpha)} \sum\limits_{k = 1}^{\alpha} H(\alpha, k) \right]}.
\end{align}
Plugging this expression into the expression for modular entropy we obtain 
\begin{align}
\tilde{S}_{\alpha}^{\rm Haar} &= -\alpha H_{\alpha-\frac{1}{2}} + \alpha H_{\alpha   +1}+\frac{1}{2} (\alpha (N-4)
   \log (2)-\log (\pi )) \notag \\
   &+\log
   \left(\frac{2^{\frac{1}{2} (-\alpha L+4 \alpha + L)} \Gamma
   \left(\alpha +\frac{1}{2}\right)}{\Gamma (\alpha+2)}\right),
\end{align}
where $H_{x} = \sum\limits   _{k = 1}^{x} \frac{1}{k}$ is are the Harmonic functions and $\Gamma (x)$ are the usual gamma functions. After taking another derivative yields closed form the expression for the capacity of entanglement
\begin{align}
   \partial_{\alpha} \tilde{S}_{\alpha}^{\rm Haar} =   \alpha \left(\Psi(\alpha+2)-\Psi
   \left(\alpha+\frac{1}{2}\right)\right),
\end{align}
where $\Psi = \partial_{\alpha} \log{(\Gamma (\alpha))}$ is the derivative of the logarithm of the Gamma function, known also as the digamma function. Evaluating this function at $\alpha = 1$ we obtain
\begin{align}
   \partial_{\alpha} \tilde{S}_{\alpha}^{\rm Haar} \Big\vert_{\alpha = 1} = \dfrac{11}{4} - \dfrac{\pi^2}{3} \approx -0.539868,  
\end{align}
to leading order in system size. Notice that this result does not explicitly depend on the system size. Numerical checks against synthetic Haar states for finite sizes (up to 20 qubits) confirm this estimate. 

\section{Ensemble relizations}\label{SMstatistics}

Throughout our work, we generate \(M\) independent realizations of the SYK-4 and SYK-2 Hamiltonians which we scale (multiply) by the scalar parameter \(g\). This method is particularly advantageous for larger system sizes, where constructing the SYK-4 Hamiltonian becomes computationally expensive and represents the main computational bottleneck. Even though the Hamiltonian is sparse, it requires significant resources to store, e.g. for $N = 32$ a single sparse realization of the SYK4 Hamiltonian takes up around 1GB of memory. The step size for the interpolation parameter used through the text \(\Delta g = 1/100\), while exclusively for the SRE computation to ease the computation load we compute it for every $\Delta g = 0.05$. The number of realizations for each fermion number $N$ is as follows:  
$N = 6$ (1000), $N = 8$ (1000), $N = 10$ (1000), $N = 12$ (400), $N = 14$ (400),  
$N = 16$ (200), $N = 18$ (200), $N = 20$ (200), $N = 22$ (100), $N = 24$ (100),  
$N = 26$ (100), $N = 28$ (50), $N = 30$ (30), and $N = 32$ (10).

\section{Analytical derivation of the logarithmic anti-flatness for the SYK-2 model}\label{app:antiflatnessderivation}

\begin{figure}[t!]
    \centering
\includegraphics[width=\columnwidth]{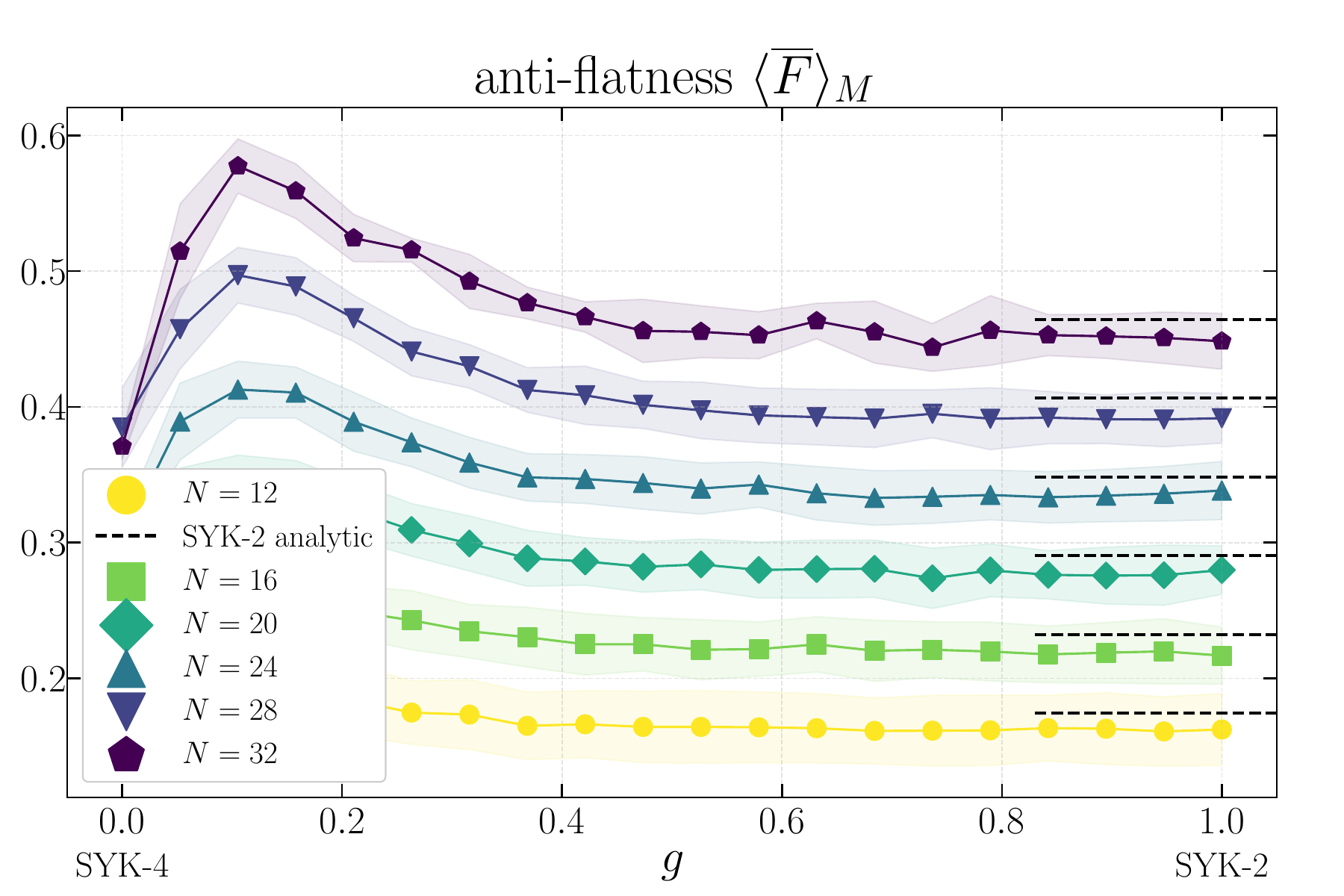}
    \caption{\justifying Ensemble averaged half-system ($f = 1/2$) anti-flatness in the ground states of the interpolated SYK model. The overline also indicates an average has been taken over the $N$ possible bipartitions.  The shaded areas represent the standard deviation across M realizations of the Hamiltonian. The dashed lines are the analytical results given in Eq.~\eqref{jednacina}.  }\label{fig-AFSM}   
\end{figure} 

 Another measure of anti-flatness that is a numerically and analytically accessible quantity is the logarithmic anti-flatness~\cite{Odavic_Viscardi_Hamma_2024}
\begin{equation}
\mathrm{F} (\rho_{R}) := 2 \left(  S_{2} (\rho_{R})  - S_{3} (\rho_{R}) \right).
\end{equation}

Here we provide some important details enabling and leading to the closed-form expression in Eq.~(14). We closely follow the derivation performed for $\alpha = 2$ R\'{e}nyi entropy in Ref.~\cite{lydzba2021entanglement, liu2018quantum}. In particular, starting from Eq.(23) in~\cite{lydzba2021entanglement} we have for the $\alpha = 3$ that 
\begin{align}
    S^{(3)}_{R} = - \dfrac{1}{2} \sum\limits_{k = 1}^{R} \ln{\left[ \lambda_k^{3} + (1 - \lambda_{k})^3 \right]}, \label{sumanchor}
\end{align}
where $\{ \lambda_{k}; \; k = 1, 2,..., R \}$ are the eigenvalues of the reduced density matrix. After tracing out degrees of freedom, the subscript $R$-th is a notation used to specify the one-body correlation (or reduced density) matrix $\rho_{R}$. The key insight in deriving $\alpha = 2$ was the observation in Ref.~\cite{liu2018quantum} that the SYK2 model subsystem eigenvalues belong to the $\beta$-Jacobi ensemble with $\beta = 2$. In particular, subsystem reduced density matrix eigenvalues, for different relative size $f$, have the following form
\begin{align}
    \mathcal{F} (f, p) = \dfrac{1}{2 \pi f} \dfrac{\sqrt{p (1 - p) + f(1 - f)- \frac{1}{4}}}{p (1 - p)},
    \label{WL}
\end{align}
where we note that this PDF is supported only on the domain $p \in [p_{-},p_{+}] $, with $p_{\pm} = \frac{1}{2} \pm \sqrt{f (1 - f)}$. The average third R\'{e}nyi entropy is obtained by
\begin{align}
    \overline{S}^{(3)} = - \dfrac{R}{2} \int {\rm d} p \mathcal{F} (f, p) \ln{\left[ p_i^{3} + (1 - p_{i})^3 \right]},
\end{align}
after replacing the sum in Eq.~\eqref{sumanchor}  as $\sum\limits_{i} \rightarrow R \int {\rm d} p F(f, p)$. We therefore have
\begin{align}
    \overline{S}^{(3)} &= -\dfrac{R}{4 \pi f} \int \dfrac{\sqrt{p (1 - p) + f(1 - f)- \frac{1}{4}}}{p (1 - p)} \notag \\
    &\times \ln{\left[ p_i^{3} + (1 - p_{i})^3 \right]}
\end{align}
Changing the variable as $p = (\lambda + 1)/2$ we obtain
\begin{align}
    \overline{S}^{(3)} = - \dfrac{R}{2 \pi f} \int\limits_{\lambda_{-}}^{\lambda_{+}} \dfrac{\sqrt{\frac{1 - \lambda^2}{4} + f(1 - f) - \frac{1}{4}}}{1 - \lambda^2} \notag \\ 
    \times \ln{\left[ \left( \dfrac{\lambda + 1}{2}\right)^3 + \left( 1 - \frac{\lambda + 1}{2} \right)^3\right]} {\rm d} \lambda,
\end{align}
where the term in the logarithm after expanding, simplifying, and rearranging reads
\begin{align}
    \ln{\left[ \left( \dfrac{\lambda + 1}{2}\right)^3 + \left( 1 - \frac{\lambda + 1}{2} \right)^3\right]}  = - \sum\limits_{n = 1}^{\infty} \frac{1}{n} \left( \frac{3}{4}\right)^n (1 - \lambda^2)^n
\end{align}
Where we expanded the logarithm as $\ln{(1 - x)} = - \sum\limits_{n = 1}^{\infty} \frac{x^n}{n}$ leading to
\begin{align}
    \overline{S}^{(3)} &= \dfrac{R}{4 \pi f} \sum\limits_{n = 1}^{\infty} \frac{1}{n} \left( \frac{3}{4}\right)^n \int\limits_{-2 \sqrt{f (1 - f)}}^{+2 \sqrt{f (1 - f)}} \left( 1 - \lambda^2 \right)^n \notag \\ &\times  \dfrac{\sqrt{ (1 - \lambda^2 ) + 4 f (1 - f) - 1}}{1 - \lambda^2}   {\rm d} \lambda.
\end{align}
The integrable can be evaluate exactly to yield $-2 (f-1) f \pi \, _2F_{1} \left( \frac{1}{2}, 1 - n, 2, -4(f-1) f \right)$. Using the following definition of the logarithmic anti-flatness
\begin{equation}
    F = 2 \left(  S_{2} (\psi_{R}) - S_{3} (\psi_{R}) \right),
\end{equation}
we trivially  obtain
    \begin{align}
        \overline{F} (R, f) =& 2 R  (1 - f) \sum\limits_{n = 1}^{\infty} \dfrac{1}{n} \left( \dfrac{1}{2^n} - \dfrac{1}{2}  \dfrac{3^n}{4^n}  \right) \,\notag \\
        &\times _2F_{1} \left( \frac{1}{2}, 1 - n\
       , 2, 4f(1 - f) \right),\label{jednacina}
    \end{align}
In Fig.~\ref{fig-AFSM} we showcase the validity of the derived expression for the SYK-2 model. For Haar random state in the large $N$ limit the logarithmic anti-flatness approach system-size-independent value of \( \mathrm{F}^{\mathrm{Haar}} = \log{(5/4)} \approx 0.223\)~\cite{Odavic_Viscardi_Hamma_2024}.

\section{Hamiltonian spectral statistics}\label{app:HamS}

\begin{figure}[h!]
    \centering
\includegraphics[width=\columnwidth]{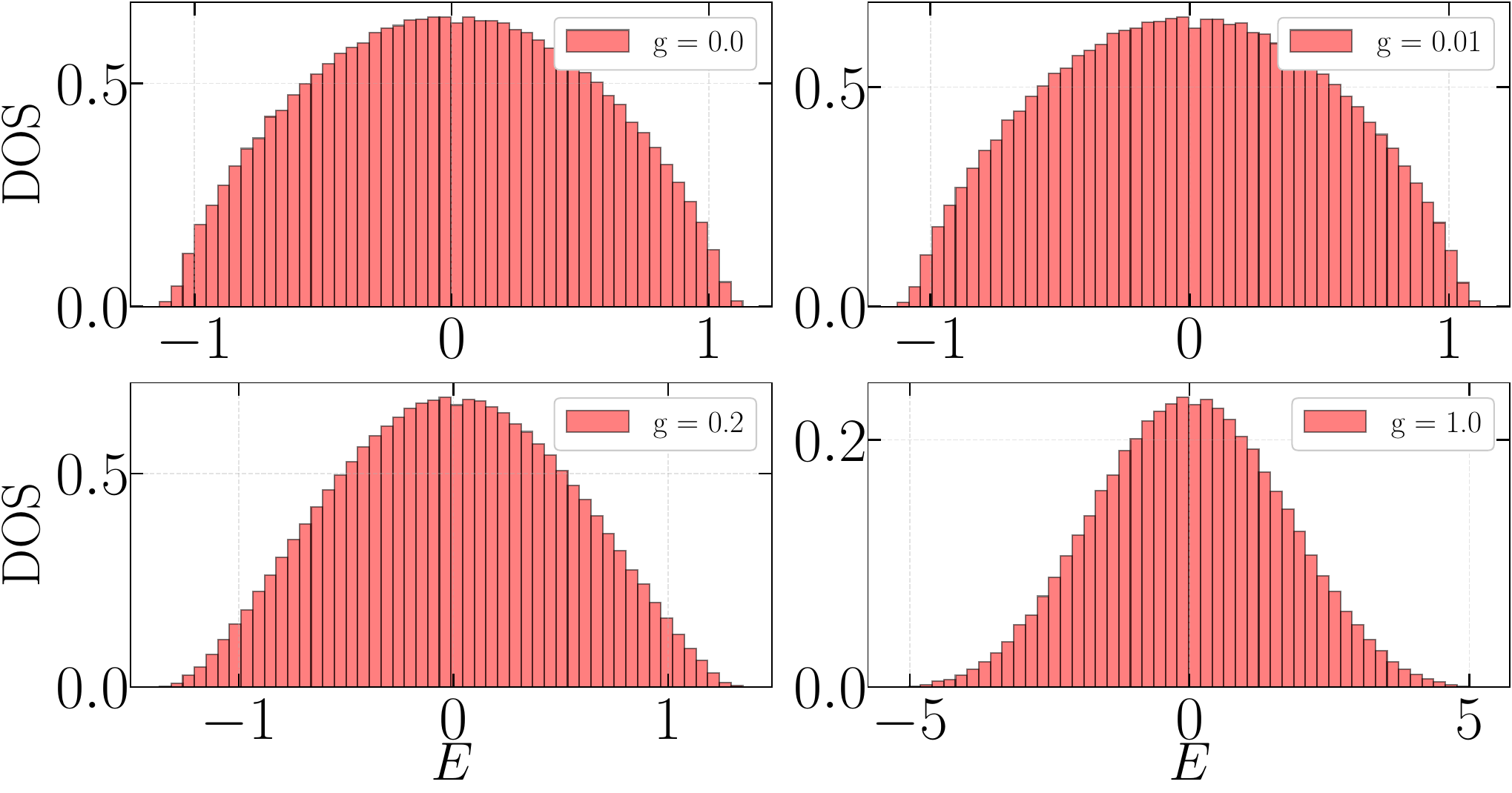}
    \caption{\justifying Density of states (DOS) of the $H_g$ spectrum for $N = 22$ fermions with support on $11$ qubits, with $M = 40$ realizations of the disordered Hamiltonian. We considered $40960$ eigenvalues for each panel and corresponding $g$. See text for details and discussions. An interesting observation regarding the DOS support is highlighted below.    }
    \label{DOS}
\end{figure}
Here we discuss the nuances related to the Hamiltonian spectrum of the interpolated model $H_g$. We first expose the results related to the full Hamiltonian spectrum of the model. More specifically, we plot the Density of states (DOS) of the spectrum in Fig.~\ref{DOS}. We can infer that its shape is akin to the one encountered in Gaussian ensembles of Random Matrix Theory (RMT) and that of Wigner semi-circle distribution. To study universal features of a system, it is essential to look at the gaps between eigenvalues rather than the eigenvalues distribution itself, that is not universal. A more detailed analysis of eigenvalues gaps of the SYK4 across different system sizes can be found in ~\cite{cotler2017black}. The SYK4 model admits different universality classes due to the existence of a particular size-dependent particle-hole symmetry of Gaussian ensembles as tabulated in table~\ref{tab:11colonne}. While the bulk spectrum of the SYK4 model does indeed behave according to RMT~\cite{garcia2018chaotic}, the spectrum at the edges is a subject of current research~\cite{altland_quantum_2024}.  
\begin{table}
    \centering
     \begin{tabular}{|>{\centering\arraybackslash}p{0.75cm}|>{\centering\arraybackslash}p{0.75cm}|>{\centering\arraybackslash}p{0.75cm}|>{\centering\arraybackslash}p{0.75cm}|>{\centering\arraybackslash}p{0.75cm}|>{\centering\arraybackslash}p{0.75cm}|>{\centering\arraybackslash}p{0.75cm}|>{\centering\arraybackslash}p{0.75cm}|>{\centering\arraybackslash}p{0.75cm}|>{\centering\arraybackslash}p{0.75cm}|}
        \hline
        N & 16 & 18 & 20 & 22 & 24 & 26 & 28 & 30   \\
        \hline
        class & GOE & GUE & GSE & GUE & GOE & GUE & GSE & GOE   \\
        \hline
    \end{tabular}
   \captionsetup{justification=centering}
   \caption{Due to a particular particle-hole symmetry, the SYK-4 model exhibits all three Gaussian ensembles~\cite{cotler2017black}.}
    \label{tab:11colonne}
\end{table}

In Fig.~\ref{DOS}, by changing the interpolation parameter $g \in [0,1]$ we infer a smooth transition from a semi-circle type DOS ($g \ll 1$) to one that of a Gaussian (as g approaches 1), final lower right panel~\cite{sarosi2017ads}. Note that we take into account every second eigenvalue since SYK4 spectrum eigenvalues are all twice degenerate. For consistency, we extend this choice of forming the DOS from every second eigenvalue for all values of $g$. Therefore the total number of eigenvalues considered for each panel of Fig.~\ref{DOS} is $M 2^{\frac{N}{2} - 1} = 40960$. 

We highlight a noteworthy aspect of the smooth change between the SYK4 and SYK2 DOS that can not be directly observed in Fig.~\ref{DOS}. More specifically, for small values of $g$ in the range $g \in [0,0.1]$, we observe the shrinking of the energy support of the Hamiltonian eigenvalues and correspondingly the DOS support. We make this observation for all system sizes up to $N= 22$ for which we computed the full spectrum. For values of $g$ that are larger than $0.1$, the support grows and the DOS starts spreading rather than shrinking. This change in behavior in the DOS could been related to our observations presented in Figs.~\ref{fig-normalizedRDM} and \ref{fig-transition} for the ground state. Interestingly, the ground state properties are sensitive to these changes in the Hamiltonian behavior, while the middle of the spectrum states are not. This phenomenology can potentially be related to the findings in~\cite{PhysRevLett.125.196602} regarding the low-temperature behavior of the interpolated SYK model.

\begin{figure}[t!]
    \begin{subfigure}{\columnwidth}
        \centering  \includegraphics[width=\linewidth]{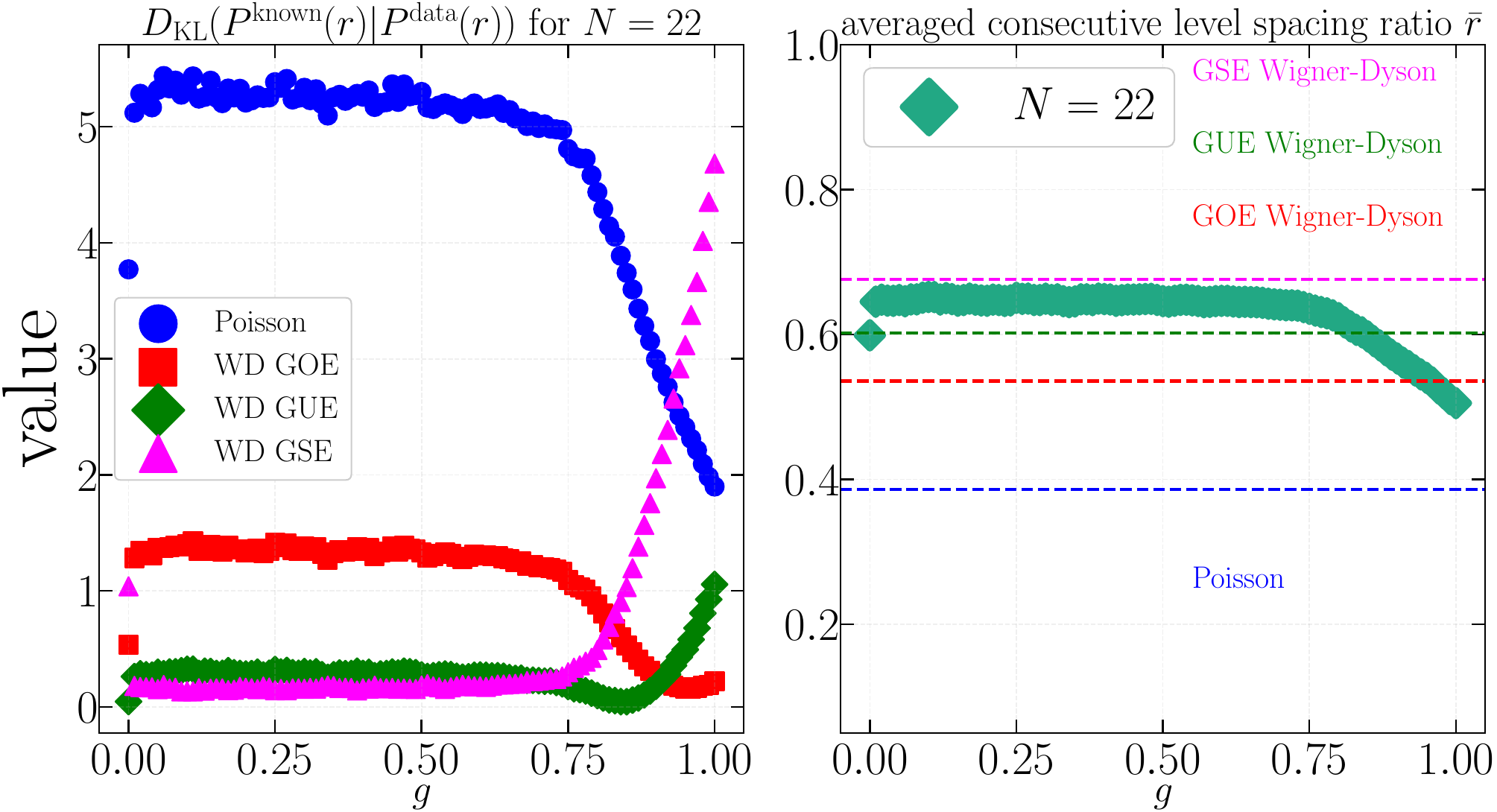}
    \end{subfigure}
    \caption{\justifying Spectral statistics results. \textit{Left panel}: KL divergence (defined in the main text) between known Gaussian ensemble RMT distributions and the underlying dataset for the consecutive gap ratios PDF $P(r)$. \textit{Right panel}: The averaged consecutive gap ratios. We considered $40960$ eigenvalues for each panel and corresponding $g$. Smaller systems size exhibit similar findings to the presented results for $N =22$ in this figure.    
     }
    \label{HamitlonianKL}
\end{figure}

To round off our short survey into the Hamiltonian spectrum we present Fig~\ref{HamitlonianKL}. In the left panel, we show the results for the KL divergence (defined in the main text) between the full Hamiltonian $H_g$ spectral statistics as compared to the well-known Gaussian RMT ensembles. The result indicates that for most values of the interpolation parameter $g$ the underlying spacing distributions are similar to the correlated Wigner-Dyson (WD) RMT ensemble expectations.  However, we note that at around $g \sim 0.75$ there is a change of behavior toward the uncorrelated Poisson distribution which we further comment on below. The explicit expressions of the distribution of consecutive gap ratios have been derived in~\cite{Atas_Bogomolny_Giraud_Roux_2013}, but for the convenience of the reader we present them below. For the standard correlated ensembles (GOE, GUE, and GSE, with $\beta = 1,2,4$ respectively) we have
\begin{align}
P^{\text{WD}}(r, \beta) = \dfrac{Z^{-1}_{\beta} \cdot (r + r^2)^\beta}{(1 + r + r^2)^{1 + \frac{3}{2} \beta}},
\end{align}
where 
\begin{align}
Z_{\beta} =
\begin{cases} 
\dfrac{8}{27}, & \beta = 1, \\[10pt]
\dfrac{4}{81} \cdot \dfrac{\pi}{\sqrt{3}}, & \beta = 2, \\[10pt]
\dfrac{4}{729} \cdot \dfrac{\pi}{\sqrt{3}}, & \beta = 4,
\end{cases}
\end{align}
while the uncorrelated ensembles typically follow the Poisson distribution, which in terms of the consecutive gap ratios reads
\begin{align}
    P^{\rm Poisson}(r) = \dfrac{1}{(1+ r)^2}.
\end{align}
We  associate consistent color coding between all figures in this work, we choose blue for the Poisson distribution, red for GOE, green for GUE and magenta for GSE.

In the right panel of Fig~\ref{HamitlonianKL} we perform a complementary analysis of the average consecutive gap ratios, defined as
\begin{align}
    \bar{r} &= \left\langle \left\langle \frac{\min (s_{k,j},s_{k+1,j})}{\max(s_{k,j},s_{k+1,j})} \right\rangle \right\rangle_{2^{N/2 - 1}-2, M} \notag \\
    &= \dfrac{1}{M (2^{N/2 - 1} - 2)} \sum\limits_{j = 1}^{M} \sum\limits_{k = 1}^{2^{N/2  - 1} - 2}  \frac{\min (s_{k,j},s_{k+1,j})}{\max(s_{k,j},s_{k+1,j})}
\end{align}
where \( s_{k,j} = \lambda_{k+1,j} - \lambda_{k,j} \), and the index \( j = 1, 2, \dots, M \) runs over $M$ available realizations while the $k$ index the available eigenvalues. It is known~\cite{Atas_Bogomolny_Giraud_Roux_2013} that the WD ratios
\begin{align}
    \bar{r}^{\rm WD-GOE} &= 4 - 2 \sqrt{3} \approx 0.53590, \\
    \bar{r}^{\rm WD-GUE} &= 2 \frac{\sqrt{3}}{\pi} - \frac{1}{2} \approx 0.60266, \\
    \bar{r}^{\rm WD-GSE} &= \frac{32}{15} \frac{\sqrt{3}}{\pi} - \frac{1}{2} \approx 0.67617,
\end{align}
while for the uncorrelated ensemble, we have
\begin{align}
    \bar{r}^{\rm Poisson} = 2 \ln{2} - 1 \approx 0.38629.
\end{align}
We observe that indeed for the system size considered ($N = 22$, but similar follows for other smaller system sizes), we recover universal statistics of the GUE for the SYK4 ($g = 0$) model that matches the GUE ensemble as previously shown in ~\cite{cotler2017black} and in agreement to Table~\ref{tab:11colonne}. Interestingly, as soon as we introduce finite $g$ the underlying distribution of the consecutive level spacing statistics changes its universality class, and more agreement is found with the GSE ensemble compared to the other distributions. This is potentially related to the fine-tuned nature of the SYK4 model and to the particle-hole symmetry that is lost at any non-vanishing $g$. Similar observation on the quantum chaotic nature of the Hamiltonian spectrum for $N \leq 22$ for all choices of the interpolation parameter has been highlighted in~\cite{garcia2018chaotic}. However, we underline that the analysis performed by the authors of~\cite{garcia2018chaotic} put the focus on the bulk spectrum, which can influence the overall value of the average gap.

Moreover, we observe that the full Hamiltonian spectrum manifestly conforms to the universal correlated statistics of RMT until around $g~\approx 0.75$ where more similarity can be found with the uncorrelated Poisson distribution (left panel). However, we note that this type of analysis is influenced by the binning of the PDF and the finite-size effects that naturally plague any finite quantum many-body system. Overall this stops us from providing conclusive judgment on the chaoticity/integrability present in the model from the full Hamiltonian spectrum. In a way, this justifies our approach to studying state complexity instead as has been done in the main text. 

In the next subsection on Entanglement Spectrum Statistics, we comment on some features that some surprising SYK model features are inherited by the eigenstates and are not restricted only to the eigenvalues.

\section{Analysis of the gap}

\begin{figure}[h!]
    \centering
\includegraphics[width=\columnwidth]{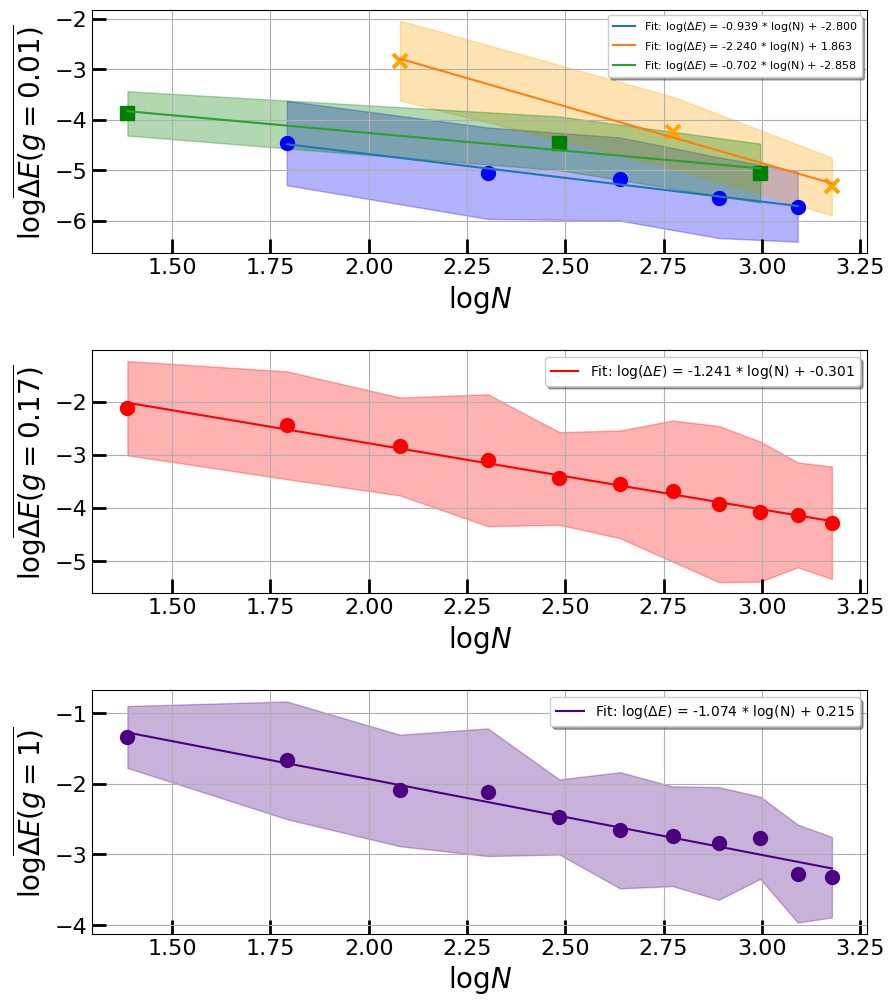}
    \caption{\justifying Power-law fits of the GSs energy gaps $\Delta E = E_1 - E_0$ of $H_g$ for different values of the interpolation parameter $g$, where $E_0$ is the GS energy and $E_1$ the first excited state energy. As soon as $g$ deviates from zero, the low-temperature density of states no longer follows an exponential distribution~\cite{chowdhury2022sachdev}. The first plot, corresponding to $g = 0.01$, shows three distinct fits, reflecting a remnant of the symmetry in the model at $g = 0$, see \cite{cotler2017black}. The orange line corresponds to GOE, the blue one to GSE and the red line to the GUE. Each data point represents an average over 100 disorder realizations for system sizes smaller than 20, and 20 disorder realizations for larger sizes.}\label{figGap}   
\end{figure} 

A comment on the behavior of the gap between the ground state and the first excited state in the interpolated SYK model is in order. In quantum many-body systems, the behavior of the energy gap is instrumental in understanding quantum phase transitions~\cite{Sachdev_2011}. Therefore, it is of value if one can infer a transition between the complex SYK4 and integrable SYK2 phase based on this commonly studied observable in quantum systems. 

The low energy spectrum of the interpolated SYK model behaves differently between the SYK4 ($g = 0$) and SYK2 $g = 1$ models.  Individual energy levels have the spacing $\sim \exp{(-N S)}$ with $S$ being the system's entropy as $N \to \infty$ for the SYK4 model, while for the SYK2 the spacing behaves as $\sim 1/N$ at the bottom of the band~\cite{chowdhury2022sachdev}. 

In Fig.~\ref{figGap}, we show the behavior of the average ground state to the first excited state gap for different choices of the interpolation parameter $g$. Fitting a power law function to the data we observe consistency with the expected $N^{-1}$ power law with exponent $-1$ as $g$ approaches 1. On the other hand, when we approach the SYK4 point, i.e. $g$ going to 0, we are still able to fit a power law function with reasonable precision. This implies that for the finite system data, it is hard to differentiate between a power-law and the exponential behavior expected for the SYK4 model, indicating the ineffectiveness of this particular probe.

\section{Entanglement Spectral Statistics (ESS) and normalized Reduced Density Matrix (RDM) spectrum}\label{app:MS}
\begin{figure}[h!]
    \centering
\includegraphics[width=\columnwidth]{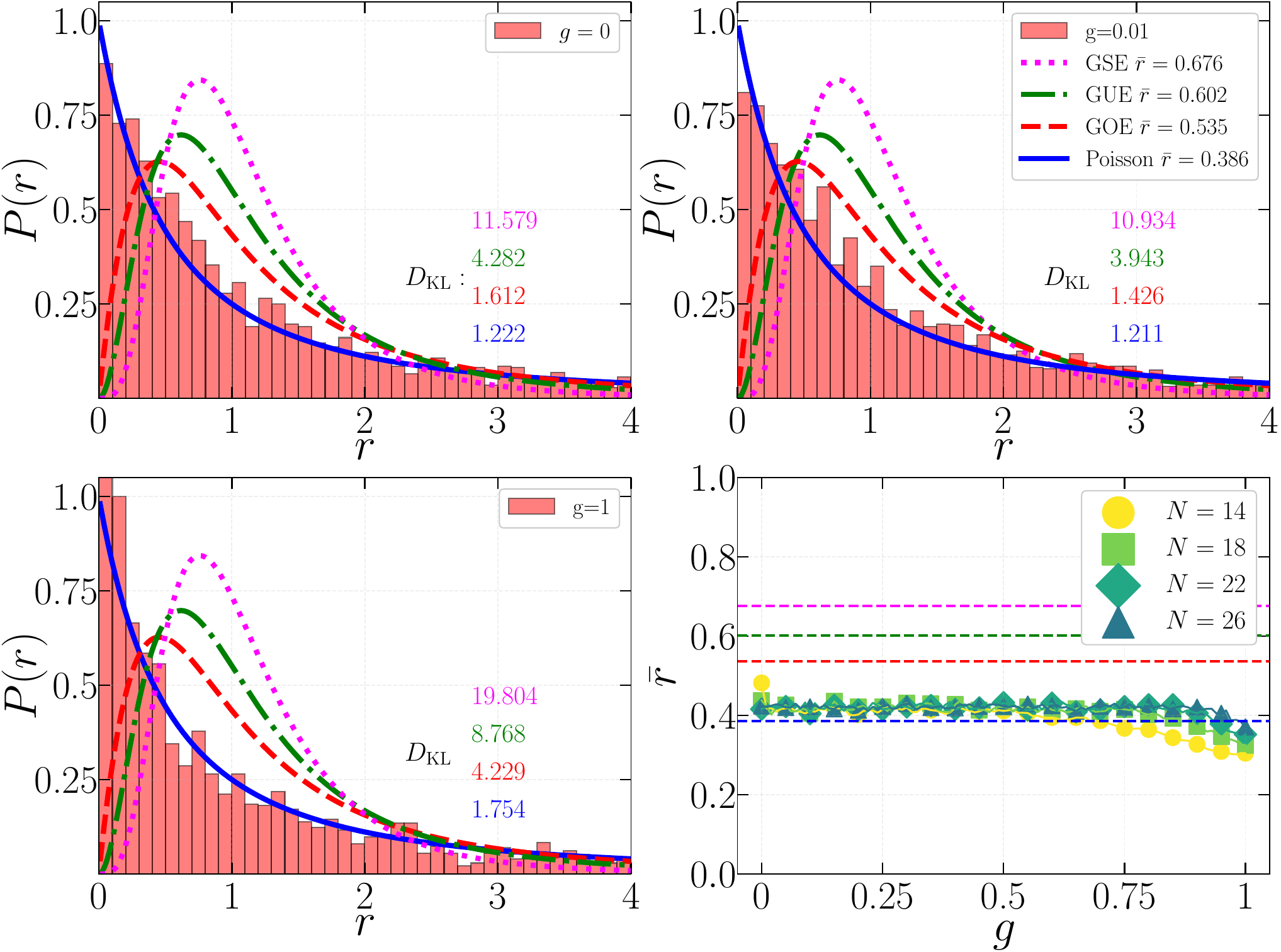}
    \caption{\justifying ESS of the middle of spectrum state RDM eigenvalues of the $H_g$ model for different values of $g$. We superimpose the analytical curves for the Wigner-Dyson (dashed) and Poisson (blue continuous) distribution for comparison. System size  $N = 22$ and number of realizations is $M = 100$. The number of bins used for the histogram is 100. The colored numbers represent the KL divergence of data against known distributions~\cite{Atas_Bogomolny_Giraud_Roux_2013} }\label{figESS_SM}   
\end{figure} 

\begin{figure}[h!]
    \begin{subfigure}{\columnwidth}
        \centering  \includegraphics[width=\linewidth]{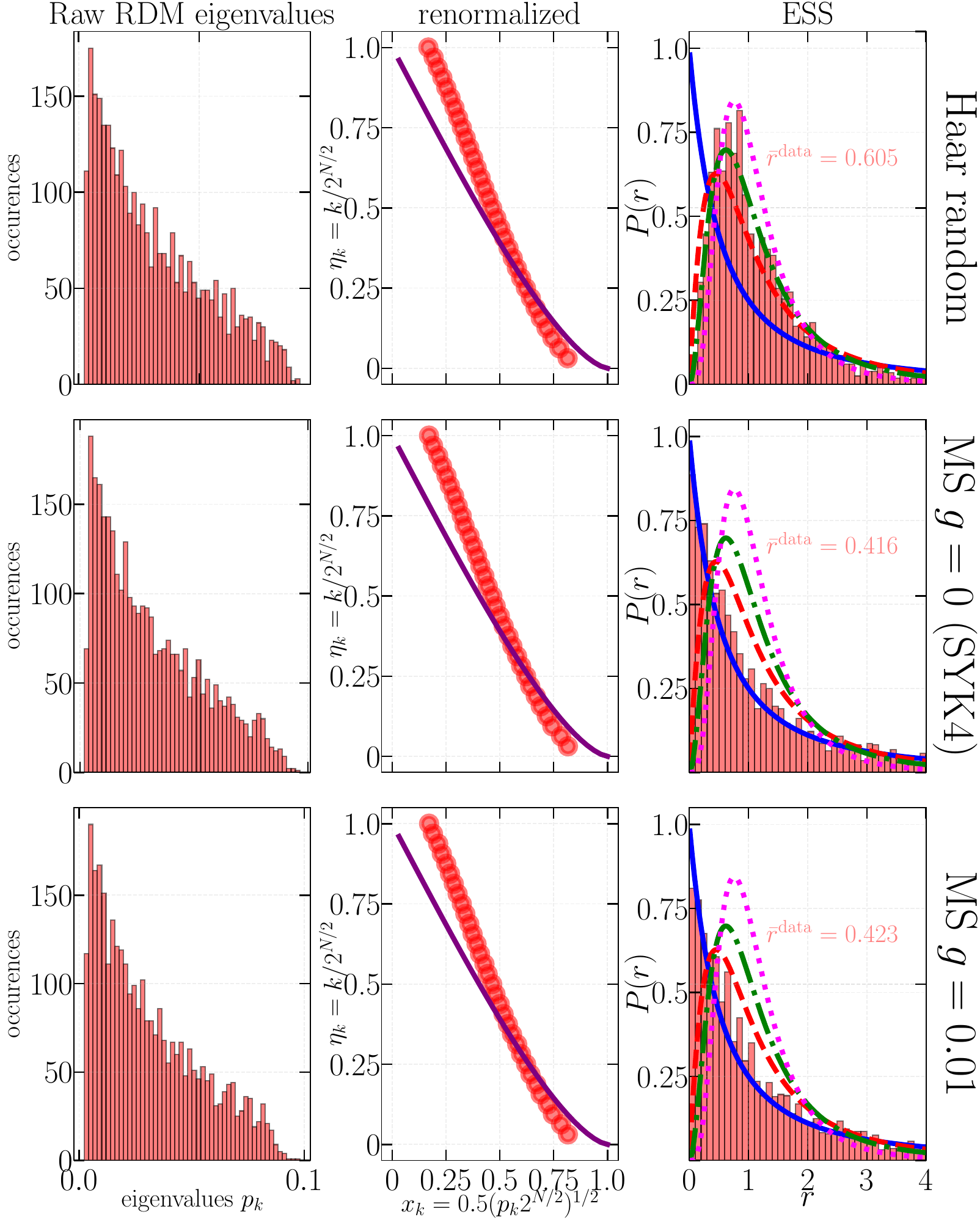}
    \end{subfigure}
    \caption{\justifying Haar vs middle of the spectrum (MS) SYK states comparative analysis across a single bipartition (same results for other bipartitions). \textit{Upper row}: Haar states included by influenced by the odd/een effects (the reason for the mismatch with the purple M-P distribution~\cite{yang2015two}); \textit{Middle row}: MS state for the $g = 0$ SYK4 Hamiltonian; \textit{Lower row}: MS state for $g = 0.01$. Other parameters: number of realization $M = 100$, nb of spins $N = 11$ (fermions $N = 22$) and across bipartition $[1,2,3,4,5]$.
     }
    \label{tot}
\end{figure}

In this section, we complete our survey of the ESS of the middle spectrum state not presented in the main text. More specifically, we repeat the same analysis presented in Fig.~(4), but for the middle of the spectrum state (MS) at $E ~\sim 0$, see Fig.~\ref{DOS}. In Fig.~\ref{figESS_SM}, we observe that for the considered system sizes the ESS does not show agreement with RMT universal ensembles and shows good agreement with the uncorrelated Poisson distribution.

In Fig.~\ref{tot}, we expand on this point and demonstrate a broader view-point of the absence of  agreement between Haar expectation in the RDM eigenvalues. In the top panel, we show the results for pure Haar random states on $N = 11$ qubits. To generate such states we simply draw real and imaginary parts of the expansion coefficients of $\vert \psi \rangle$ as random Gaussian variables, and normalize the output vector state~\cite{zyczkowski2000truncations}. This procedure provides an excellent agreement with the predicted Marchenko-Pasture (M-P) distribution defined in the main text and represented with the purple line in the middle panels. We note that the disagreement between obtained values and the M-P distribution comes from the odd/even effect that vanishes in the thermodynamic limit. The final panel of the upper row shows that ESS for the Haar states does conform with the RMT prediction and universal GUE ensemble. The Haar example serves as a reference point to the results we obtain for the MS state of the interpolated model. More specifically, we can observe that the RDM spectrum itself shows quite similar features to that of a Haar random state, however, the ESS point to an uncorrelated ensemble and the non-universal Poisson distribution. These unorthodox features of SYK4 and of the interpolated model, when it comes to universality statistics probes, are most likely part of the broader context of the particular structure that the eigenstates entail and are evidenced already in the above mentioned section on the Hamiltonian spectrum statistics.

\section{Stabilizer R\'{e}nyi Entropy of Middle of the spectrum states}\label{app:SREMS}
\begin{figure}[h!]
    \centering
\includegraphics[width=\columnwidth]{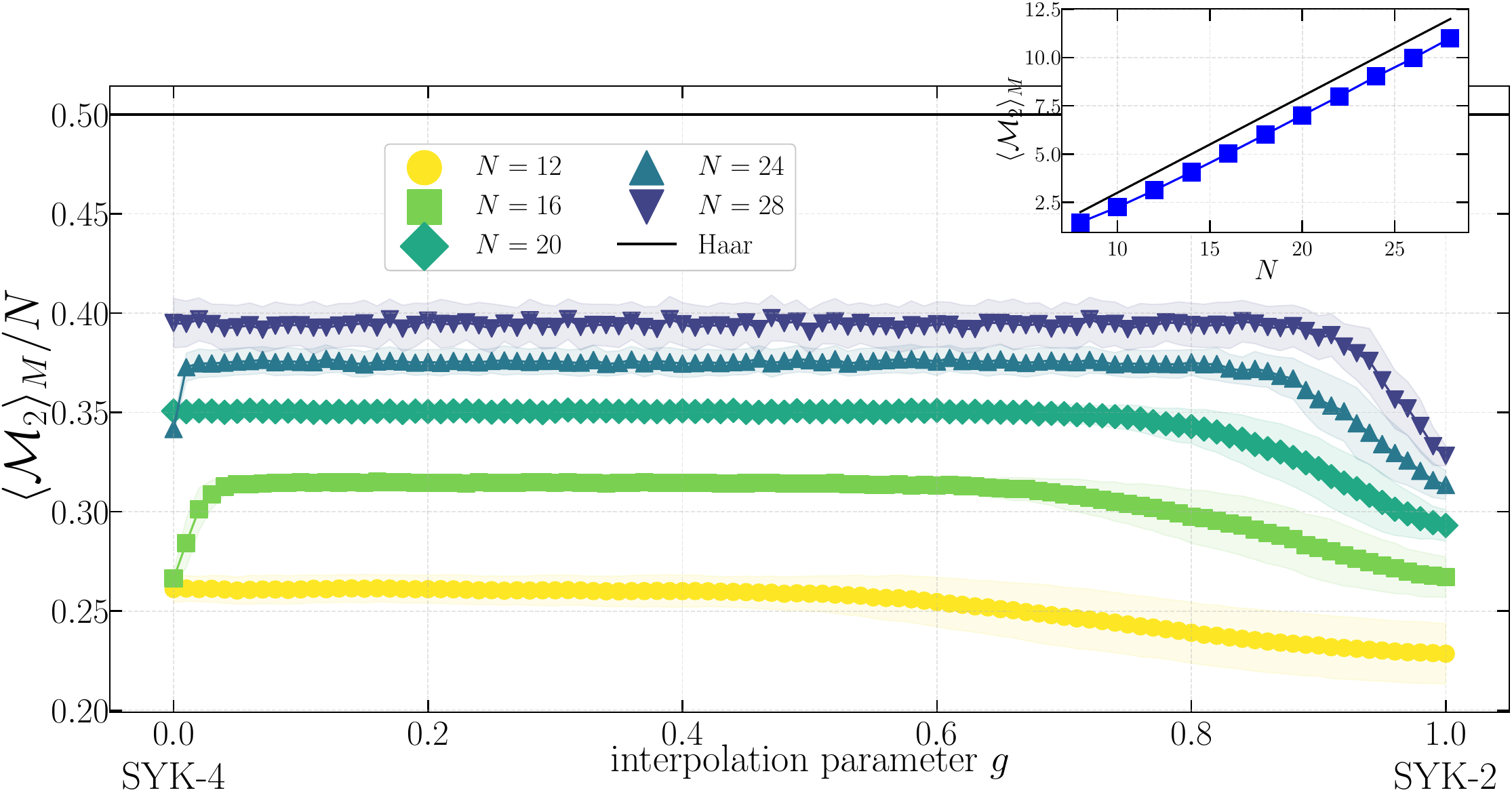}
    \caption{\justifying Ensemble-averaged SRE $\langle \mathcal{M}_{2} \rangle_{M}$ in middle-of-the-spectrum (MS) states of the interpolated Hamiltonian $H(g)$. \textit{Inset:} total SRE for a fixed interpolation parameter $g = 0.3$ as a function of system size $N$.
}\label{figMiddleSRE}   
\end{figure}

We present results for the Stabilizer R\'enyi entropy (SRE) measured in middle-of-the-spectrum eigenstates of the interpolated SYK Hamiltonian $H(g)$. As shown in Fig.~\ref{figMiddleSRE}, the behavior of the SRE qualitatively resembles that of the entanglement entropy for these states: the SYK-4 phase dominates across most values of the interpolation parameter $g$. In the inset of the figure, we show how the SRE changes with system size for value $g = 0.3$, which is deep inside the SYK-4 regime and does not allow for symmetry constraints present for the pure $g=0$ SYK-4 model. In particular, in terms of non-stabilizerness, also high-energy states fail to reach Haar-type universality.

\end{document}